\documentclass[10pt,onecolumn]{IEEEtran}
\usepackage{amsmath}
\usepackage{amsthm}
\usepackage{amssymb}
\usepackage{graphicx}
\usepackage{multirow}

\usepackage{subfig}

\newcommand{\abs}[1]{|#1|}

\newcommand{\labs}[1]{\|{#1}\|_{1}}
\newcommand{\defn}{\triangleq}
\newcommand{\eu}[1]{\|{#1}\|_{2}}

\newtheorem{theorem}{Theorem}
\newtheorem{lemma}{Lemma}
\newtheorem{cor}{Corollary}
\newtheorem{define}{Definition}
\newtheorem{alg}{Algorithm}

\begin{document}

\title{Detectability of Symbol Manipulation by an Amplify-and-Forward
  Relay \thanks{This paper was presented in part at IEEE INFOCOM 2012.}}

\author{Eric Graves and Tan F. Wong \\
  Department of Electrical \& Computer Engineering \\
  University of Florida, FL 32611 \\
\texttt{\{ericsgra,twong\}@ufl.edu}}

\maketitle

\begin{abstract}
  This paper studies the problem of detecting a potential malicious
  relay node by a source node that relies on the relay to forward
  information to other nodes. The channel model of two source nodes
  simultaneously sending symbols to a relay is considered. The relay
  is contracted to forward the symbols that it receives back to the
  sources in the amplify-and-forward manner. However there is a chance
  that the relay may send altered symbols back to the sources. Each
  source attempts to individually detect such malicious acts of the
  relay by comparing the empirical distribution of the symbols that it
  receives from the relay conditioned on its own transmitted symbols
  with known stochastic characteristics of the channel. It is shown
  that maliciousness of the relay can be asymptotically detected with
  sufficient channel observations if and only if the channel satisfies
  a non-manipulable condition, which can be easily checked.  As a
  result, the non-manipulable condition provides us a clear-cut
  criterion to determine the detectability of the aforementioned class
  of symbol manipulation attacks potentially conducted by the relay.
\end{abstract}
\begin{IEEEkeywords}
Maliciousness detection, symbol manipulation, amplify-and-forward
relay, physical-layer security, trust metric valuation.
\end{IEEEkeywords}

\IEEEpeerreviewmaketitle

\section{Introduction}
\label{sec:intro}

It is a commonplace in many communication networks that no direct
physical link exists between the source and destination nodes of an
information flow. Thus information needs to be relayed from the source
to the destination through intermediate nodes. This requirement brings
forth a major security question: How is one able to ensure that the
intermediate nodes faithfully forward information from the source to
the destination? Trust management
\cite{Blaze1996}\nocite{Hubaux01}--\cite{Buttyan07} is a widely
researched approach to address this question. In essence, trust
management pertains to the establishment, distribution, and
maintenance of trust relationships among nodes in a network. Based
upon such relationships, it is expected that trusted nodes will
faithfully operate according to some protocols that they have agreed
upon.

The aforementioned trust relationships are primarily quantified using
trust metrics that are evaluated through nodes interacting with and
observing the behaviors of each other
\cite{Trust}\nocite{MalCoop,TrustVector,Game,JiangB05}--\cite{Theo04}. For
nodes that do not directly interact with each other, trust
relationships can be established and maintained via inference
\cite{ZhangInfocom10}. It is clear that the valuation of trust metrics
is critically important in this trust management approach. Many
different ways have been proposed to evaluate the trust metrics based
on authentication keys \cite{Capkun03,Beth94,ZhangInfocom08},
reputation \cite{Sun06,Oli08}, and evidence collected from network as
well as physical interaction
\cite{Josang99,Theo06,CapkunMobihoc03}. Given the complexity and
difficulty involved in quantifying the vague notion of trust, one
would expect these valuation schemes are naturally ad hoc.

To more systematically construct a trust metric, one needs to specify
the class of malicious actions against which the metric measures.  In
this paper, we consider a class of data manipulation attacks in which
intermediate nodes may alter the channel symbols that they are
supposed to forward.  We cast the trust metric valuation problem as a
maliciousness detection problem against this class of attacks. More
precisely, a node (or another trusted node called a \emph{watchdog}
\cite{Marti00}) detects if another node relays manipulated symbols
that are different from those originally sent out by the node itself.
Observations for detection against the malicious attack can be made in
the physical and/or higher layers.

Most existing maliciousness detection methods are key-based, requiring
at the minimum the source and destination nodes to share a secret key
that is not privy to the relay node being examined.  In
\cite{NullKeys}, keys, which correspond to vectors in a null space,
are given to all nodes in a network. Any modification to the encrypted
data by a relay node can be checked by determining whether the
observed data falls into the null space or not. In \cite{Ariadne},
symmetric cryptographic keys are applied to the data at source and
destination for the purpose of checking the maliciousness of relay
node(s).  Another key-based approach is considered in
\cite{Papadimitratos02} by measuring whether only a small amount of
packets are dropped by relay nodes.  In \cite{Mao}, a cross-layer
approach based on measurement of channel symbols is taken.  Two keys
are employed in that scheme; one to create a set of known data and
another to make the data indistinguishable from the key. When the
destination receives the message, the probability of error of the
transmitted values of the key can be used to determine if the relay
node is acting maliciously. In all, the key-based maliciousness
detection schemes described above are far from desirable as they
require the support of some key distribution mechanism, which in turn
presumes the existence of inherently trusted nodes in the network.
Moreover some key-based methods have been shown insecure in
\cite{Mitchell01} when nondeterminism and bit-level representation of
the data is considered.

For the class of symbol manipulation attacks, it is intuitive that
maliciousness of a node should be detected by the nearby nodes based
on measurements obtained at the lower layers, since such measurements
are more reliable than those made by faraway nodes and at the higher
layers as there are fewer chances for potential adversaries to tamper
with the former measurements.  Hence we investigate the maliciousness
detectability problem from a physical-layer perspective by considering
a model in which two sources want to share information through a
potentially untrustworthy relay node that is supposed to relay the
information in the \emph{amplify-and-forward} manner. In
\cite{GravesInfocom12}, we provide a preliminary study on the problem
under a restrictive case in which the relay may only modify the
channel symbols based on some independent and identically distributed
(i.i.d) attack model, and the source nodes can perfectly observe the
symbols forwarded by the relay. In this paper, we extend the treatment
to a general channel model and an effectively general class of symbol
manipulation attacks.  The details of the channel and attack models
are provided in Section~\ref{MandA}.

In Section~\ref{se:motivate}, we qualitatively discuss maliciousness
detectability in a simple binary-input addition channel in order to
motivate the detectability problem. This example has been presented
in~\cite{GravesInfocom12}. It is repeated here for easy reference.  In
Section~\ref{sec:main}, we state our main result, which is a necessary
and sufficient condition on the channel that guarantees asymptotic
detection of maliciousness individually by both source nodes using
empirical distributions of their respective observations.  We also
provide algorithms to check for the stated condition. The results
presented in Section~\ref{sec:main} make clear that maliciousness
detectability under our model is a consequence of the stochastic
characteristics of the channel and the sources. It works solely based
on observations made by the source nodes about the symbols sent by the
relay node together with knowledge about the channel.  No presumed
shared secret between any set of nodes is required or used. Thus the
proposed maliciousness detection approach can be used independent of
or in conjunction with the key-based methods described above to
provide another level of protection against adversaries

The proofs of the results in Section~\ref{sec:main} are provided later
in Section~\ref{sec:proofs}. In Section~\ref{sec:examples}, we present
results from numerical simulation studies of the addition channel
example in Section~\ref{se:motivate} and other more complicated
channels to illustrate the asymptotic detectability results in
Section~\ref{sec:main} with finite observations.  Finally we draw a
few conclusions about this work in Section~\ref{sec:conclusion}.

\section{Motivating Example}
\label{se:motivate}
\begin{figure}
\centering
\includegraphics[width=0.45\textwidth]{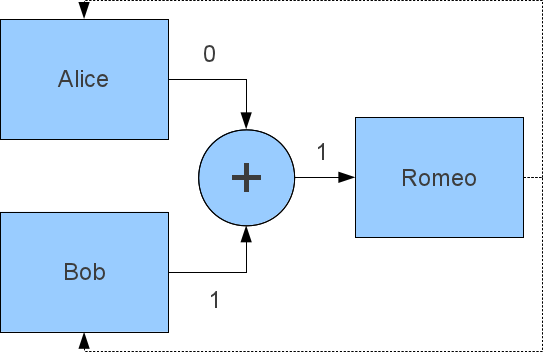}
\caption{Motivating example of a binary-input addition channel.}
\label{fig:motivate}
\end{figure}

To motivate the maliciousness detectability model and results in later
sections, let us first consider the simple binary-input addition
channel shown in Fig.~\ref{fig:motivate}, in which two source nodes
(Alice \& Bob) communicate to one another through a relay node (Romeo)
in discrete time instants.
The source alphabets of Alice and Bob are both binary $\{0,1\}$. The
channel from Alice and Bob to Romeo is defined by the summation of the
symbols transmitted by Alice and Bob. Romeo is supposed to broadcast
his observed symbol, without modification, back to Alice and Bob. Both
Alice and Bob observe the symbol transmitted by Romeo perfectly. Thus
the input and output alphabets of Romeo and the observation alphabets
of both Alice and Bob are all ternary $\{0,1,2\}$. Alice, for
instance, can obtain Bob's source symbol by subtracting her own source
symbol from the symbol transmitted by Romeo.

Now consider the possibility that Romeo may not faithfully forward the
symbol that he observes to Alice and Bob in an attempt to impede the
communication between them. The main question that we are interested
in is whether Alice and Bob are able to discern, from their respective
observed symbols, if Romeo is acting maliciously by forwarding symbols
that are different than those he has received.  To proceed answering
this question, let us first consider a single round of transmission,
i.e., Alice and Bob transmit their source symbols to Romeo and then
Romeo broadcasts a symbol (may be different than what he has received)
back to Alice and Bob. Suppose that Alice sends a $0$ and receives a
$2$ back from Romeo. Then it will be clear to her that Romeo must have
modified what he has received. On the other hand, if Alice receives a
$1$ back, then she will not be able to tell whether Romeo has acted
maliciously or not. One can continue this line of simple deduction to
obtain all the possible outcomes in Table~\ref{tb:motivate_outcome}.
\begin{table}
\centering
\caption{Possible outcomes of the system in
  Fig.~\ref{fig:motivate}.}
\label{tb:motivate_outcome}
\begin{tabular}{|c|c|c|c|c|}
\hline
\multirow{2}{*}{Alice} & \multirow{2}{*}{Bob} & Romeo& Romeo&
\multirow{2}{*}{Detection Outcome}\\
 & & In&  Out& \\
\hline
\multirow{3}{*}{0} & \multirow{3}{*}{0} & \multirow{3}{*}{0} & 0 & \textsl{Not malicious} \\
\cline{4-5}
& & & 1 &  \textbf{Not detected} \\
\cline{4-5}
& & & 2 & Alice \& Bob both detect \\
\hline
\multirow{3}{*}{0} & \multirow{3}{*}{1} & \multirow{3}{*}{1} & 0 & Bob
detects \\
\cline{4-5}
& & & 1 & \textsl{Not malicious} \\
\cline{4-5}
& & & 2 & Alice detects \\
\hline
\multirow{3}{*}{1} & \multirow{3}{*}{0} & \multirow{3}{*}{1} & 0 &
Alice detects \\
\cline{4-5}
& & & 1 & \textsl{Not malicious} \\
\cline{4-5}
& & & 2 & Bob detects \\
\hline
\multirow{3}{*}{1} & \multirow{3}{*}{1} & \multirow{3}{*}{2} & 0 &
Alice \& Bob both detect \\
\cline{4-5}
& & & 1 & \textbf{Not detected} \\
\cline{4-5}
& & & 2 & \textsl{Not malicious} \\
\hline
\end{tabular}
\end{table}
It is clear from the table that neither Alice nor Bob will be able to
determine if Romeo is malicious in general from a single round of
transmission.

However the situation changes if Alice and Bob know the source
distributions of one another and are allowed to decide on the
maliciousness of Romeo over multiple rounds of transmission.  To
further elaborate, suppose that the source symbols of Bob and Alice
are i.i.d. Bernoulli random variables with parameter
$\frac{1}{2}$. Then the probabilities of the events that Romeo's input
symbol takes on the values $0$, $1$, and $2$ are $\frac{1}{4}$,
$\frac{1}{2}$, and $\frac{1}{4}$, respectively. In particular, out of
many rounds of transmission one would expect half of the symbols
transmitted by Romeo be $1$'s.  From Table~\ref{tb:motivate_outcome},
we see that for Romeo to be malicious and remained undetected by
neither Alice nor Bob, he can only change a $0$ to a $1$ and a $2$ to
a $1$ in any single round of transmission. But if he does so often,
the number of $1$'s that he sends out will be more than half of the
number of transmission rounds, as expected from normal operation.  On
the other hand, Romeo can fool one of Alice and Bob by changing a $1$
to either a $0$ or $2$. But he is not able to determine in each change
whether Alice or Bob is fooled. Hence over many such changes the
probability of not being detected become decreasingly small.  In
summary, Alice and Bob may individually deduce any maliciousness of
Romeo by observing the distribution of Romeo's output symbol
conditioned on their respective own input symbols. This capability is
induced by the restrictions on what Romeo can do that are imposed by
the characteristic of the addition channel depicted in
Fig.~\ref{fig:motivate}. It is important to notice that Alice and Bob
do not need to possess any shared secret that is not privy to Romeo.

\section{ System Model}\label{MandA}

\subsection{Notation}
Let $a$ be a $1\!\times\!m$ row vector and $A$ be a $m\!\times\! n$
matrix. For $i=1,2,\ldots,m$ and $j=1,2,\ldots,n$, $a_i$ and $[a]_i$
both denote the $i$th element of $a$, and $A_{i,j}$ and $[A]_{i,j}$
both denote the $(i,j)$th element of $A$. Whenever there is no
ambiguity, we will employ the unbracketed notation for
simplicity. Moreover, we write the $i$th row of $A$ as $A_i$. Of
course, $[A_i]_j = A_{i,j}$. Let the transpose of $A$ be denoted by
$A^T$. Then the $j$th column of $A$ is $(A^T_j)^T$.  It is our
convention that all column vectors are written as the transposes of
row vectors. For instance, $a$ is a row vector and $a^T$ is a column
vector.  The $L_1$-norm of $a$ is $\labs{a} = \sum_{i=1}^{m}
\abs{a_i}$, while the Euclidean norm of $a$ is $\eu{a} = \sqrt{a
  a^T}$. The operation $\mathrm{vec}(A)$ vectorizes the matrix $A$ by
stacking its columns to form a $mn\!\times\!1$ column vector. We
define $\labs{A} \defn \labs{\mathrm{vec}(A)}$ and $\eu{A} \defn 
\eu{\mathrm{vec}(A)}$. Note that $\eu{A}$ is the Frobenius norm of
$A$.  The identity and zero matrices of any dimension are denoted by
the generic symbols $I$ and $0$, respectively.

Let $X$ be a discrete random variable. We use $\abs{X}$ to denote the
size of the alphabet of $X$. Our convention is to use the
corresponding lowercase letter to denote the elements in the alphabet
of a random variable. For example, the alphabet of $X$ is
$\{x_{1},x_{2}, \ldots, x_{\abs{X}}\}$. We denote the probability mass
function (pmf) $\Pr(X=x_j)$ by $p(x_j)$. In addition, let $X$ and $Y$
be two discrete random variables. We denote the conditional pmf
$\Pr(Y=y_i|X=x_j)$ by $p(y_i|x_j)$ for simplicity. Let $x^N$ denote a
sequence of $N$ symbols drawn from the alphabet of $X$. The counting
function $\pi(x_i;x^N)$ denotes the number of occurrences of $x_i$,
the $i$th symbol in the alphabet of $X$ as described above, in the
sequence $x^N$. Let $1_n(x_i;x^N)$ be the indicator function of the
condition that the $n$th symbol in the sequence $x^N$ is $x_i$. Then
we clearly have
\[
\pi(x_i;x^N) = \sum_{n=1}^{N} 1_n(x_i;x^N).
\]
The counting function also trivially extends to give the number of
occurrences of a tuple of symbols drawn from the corresponding tuple
of alphabets of random variables. For example, if $x^N$ and $y^N$ are
length-$N$ sequences of symbols drawn from the alphabets of $X$ and
$Y$, respectively, then
\[
\pi(x_i,y_j;x^N,y^N) = \sum_{n=1}^{N} 1_n(x_i;x^N) 1_n(y_j;y^N).
\]
The set of typical $X$-sequences (e.g., \cite[Definition
6.1]{Yeung2008}) is denoted by
\[
T^N_{[X],\delta} \defn \left\{ x^N: \sum_{i=1}^{\abs{X}} \left| 
    \frac{1}{N}\pi(x_i;x^N) - p(x_i) \right| \leq \delta \right\}.
\]
Whenever there is no confusion, we write for instance $\pi^N(x_i)$ and
$1_n^N(x_i)$ in place of $\pi(x_i;x^N)$ and $1_n(x_i;x^N)$,
respectively, to simplify notation. Finally, let us define the symbol
indexing maps $\chi_n(x^N)$, $n=1,2,\ldots,N$, by assigning the index
value $j$ to $\chi_n(x^N)$ when the $n$th symbol in the sequence $x^N$
is $x_j$, namely, the $j$th element in the alphabet of $X$.

\subsection{Channel model} \label{sec:model}

\begin{figure}
  \centering 
\subfloat[Multiple-access channel (time instants
  $1,2\ldots,N$)]{\includegraphics[width=0.475\textwidth]{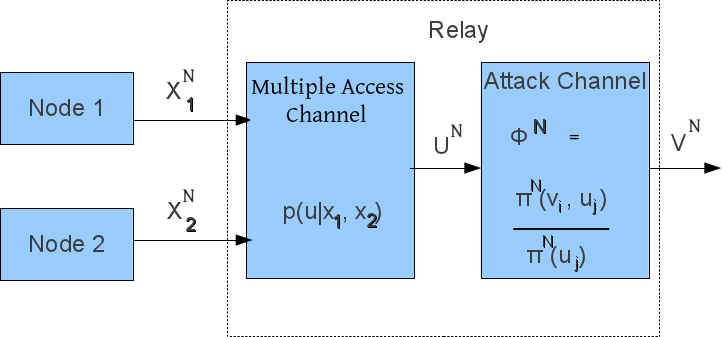}\label{fig:model1}}

\subfloat[Broadcast channel (time instants
  $N+1,N+2,\ldots,2N$)]{\includegraphics[width=0.465\textwidth]{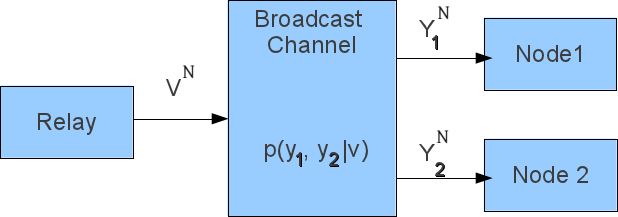} \label{fig:model2}}
\caption{Amplify-and-forward relaying model.}
\label{fig:model}
\end{figure}

Consider the channel model shown in Fig.~\ref{fig:model}. This model
serves as a generalization of the motivating example described in
Section~\ref{sec:intro}.
Two nodes (1 and 2) simultaneously forward their source symbols to a
relay node.  The relay node is supposed to forward its received
symbols back to the two nodes (or to some other nodes) in the
amplify-and-forward manner. There is some possibility that the relay
may modify its received symbols in an attempt to degrade the
performance of the transmission. Our goal is to determine if and when
it is possible for nodes 1 and 2 to detect any malicious act of the
relay by observing the symbols broadcast from the relay in relation to
the symbols that they individually transmitted. Note that the above
model also covers the perhaps more common scenario in which only one
node has information to send while the transmission by the other node
is regarded as intentional interference.

More specifically, let $X_1$ and $X_2$ be two independent discrete
random variables that specify the generic distributions of the symbols
transmitted by nodes 1 and 2, respectively. At time instants
$1,2,\ldots,N$, nodes 1 and 2 transmit i.i.d. symbols respectively
distributed according to $X_1$ and $X_2$. The transmission goes
through a memoryless multiple-access channel (MAC) with the random
variable $U$ describing its generic output symbol. The MAC is
specified by the conditional pmf $p(u_k|x_{1,i}, x_{2,j})$.  The relay
node, during time instants $1,2,\ldots,N$, observes the output symbols
of the MAC, processes (or manipulates) them, and then broadcasts the
processed symbols out to nodes 1 and 2 at time instants
$N+1,N+2,\ldots,2N$ via a memoryless broadcast channel (BC) with the
random variables $V$ describing its generic input symbol and $Y_1$ and
$Y_2$ describing the generic output symbols at nodes 1 and 2,
respectively. The BC is specified by the conditional pmf
$p(y_{1,i},y_{2,j}|v_k)$. In addition, because the relay is supposed
to work in the amplify-and-forward manner, we adopt the reasonable
model that the alphabets of $U$ and $V$ are of the same size, and
there is a one-to-one correspondence $u_i \leftrightarrow v_i$,
$i=1,2,\ldots,\abs{U}$, between elements of the alphabets.

Let $u^N$ denote the sequence of $N$ output symbols of the MAC
observed by the relay during time instants $1, 2, \ldots, N$, and
$v^N$ denote the sequence of $N$ input symbols of the BC transmitted
by the relay during time instants $N+1,N+2,\ldots,2N$. Then the
mapping $v^N = \phi^N (u^N)$ represents the manipulation performed by
the relay. The \emph{manipulation map} $\phi^N$ is allowed to be
arbitrary, deterministic or random, and known to neither node~1
nor~2. The only restriction we impose is the Markovity condition that
$p(v^N|u^N,x_1^N, x_2^N) = p(v^N|u^N)$, where $x_1^N$ and $x_2^N$
denote the symbol sequences transmitted by nodes 1 and 2,
respectively, during time instants $1,2,\ldots,N$. That is, the relay
may potentially manipulate the transmission based only on the output
symbols of the MAC that it observes.

\subsection{Maliciousness of relay}
Consider the $\abs{U}\!\times\!\abs{U}$ matrix $\Phi^N$ whose
$(i,j)$th element is defined by
\begin{equation} \label{eq:attack}
\Phi^{N}_{i,j} \defn \frac{\pi^N(v_{i},u_{j})}{\pi^N(u_{j})}.
\end{equation}
It is obvious that $\Phi^N$ is a stochastic matrix describing a valid
conditional pmf of a fictitious channel, which we will refer to as the
\emph{attack channel}. Despite omitted from its notation, the attack
channel $\Phi^{N}$ depends on the sequences $u^N$ and $v^N$. Rather
than directly acting on the MAC output symbols to produce input
symbols for the BC, the attack channel $\Phi^N$ extracts the
statistical properties of the manipulation map $\phi^N$ that are
relevant to our purpose of defining (and later detecting) the
maliciousness of the action of the relay:
\begin{define} \label{def:maliciousness} \textbf{(Maliciousness)}
  The relay is said to be non-malicious if $\labs{\Phi^N - I}
  \rightarrow 0$ in probability as $N$ approaches infinity. Otherwise,
  the relay is considered malicious.
\end{define}
In the strictest sense, normal amplify-and-forward relay operation
should require $\Phi^N = I$ for all $u^N$ and $v^N$ and for each $N$.
Nevertheless it turns out to be beneficial to consider the relaxation
in Definition~\ref{def:maliciousness} when the primary focus is to
check whether the relay is degrading the channel rather than attacking
a specific part of the transmission. In particular, the probabilistic
and limiting relaxation in Definition~\ref{def:maliciousness} allows
us to obtain definite results (see Section~\ref{sec:main}) for the
very general class of potential manipulation maps described above by
tolerating actions, such as manipulating only a negligible fraction of
symbols, that have essentially no effect on the information rate
across the relay. We point out that it is possible to develop similar
results based on the above-mentioned strictest sense of maliciousness
for some more restricted classes of manipulation maps (see
\cite{GravesInfocom12} for instance).

\subsection{Maliciousness detection} \label{sec:attackdetect}
Due to symmetry, it suffices to focus on node 1's attempt to detect
whether the relay is acting maliciously or not. To that end, we are
particularly interested in the ``marginalized'' MAC pmf
$p(u_i|x_{1,j})$ and marginal BC pmf $p(y_{1,i}|v_j)$. For better
bookkeeping, we will write the two conditional pmfs in terms of the
$\abs{U}\!\times\!\abs{X_1}$ matrix $A$ and
$\abs{Y_1}\!\times\!\abs{U}$ matrix $B$ whose elements are
respectively defined by
\begin{align*}
  A_{i,j} &\defn p(u_{i} \vert x_{1,j}) \\
  B_{i,j} &\defn p(y_{1,i} \vert v_{j}).
\end{align*}
To complete the bookkeeping process, we define the
$\abs{Y_1}\!\times\!\abs{X_1}$ matrix $\Gamma^N$ as
\begin{equation}\label{e:gammaN}
  \Gamma^N \triangleq  B \Phi^N A,
\end{equation}
which can be interpreted as the conditional pmf of the node 1's
observation if the relay were to act in an i.i.d. manner described by
the attack channel $\Phi^N$.

We assume that node 1 knows $A$ and $B$. We will refer to the pair
$(A,B)$ as the \emph{observation channel} for node~1, which may use
knowledge about the observation channel to detect any maliciousness of
the relay. Justifications for this assumption can be made based on
applying knowledge of the physical MAC and BC in a game-theoretic
argument similar to the one given in \cite{Mao}. Before data
communication between the nodes and relay takes place, they must agree
on a relaying protocol. During the negotiation process of such
protocol, the relay needs to either reveal $A$ and $B$ to the nodes,
or provide assistant to the nodes to learn $A$ and $B$.  If the relay
provides false information about $A$ and $B$ corresponding to a more
favorable channel environment than the actual one, the nodes may use
the knowledge of the physical channel to check against the false
information. On the other hand, it would also not be beneficial for
the relay to misinform the nodes with a less favorable channel
environment, since in such case the nodes may simply decide not to use
the relay.

We may also assume that $A$ contains no all-zero rows and that
$p(x_{1,j}) > 0$ for all $x_{1,j}$ in the alphabet of $X_1$. If these
assumptions do not hold, we can reinforce them by removing symbols
from the alphabets of $X_1$ and $U$ and deleting the corresponding
rows and columns from $A$, without affecting the system model. In
addition, note that relabeling of elements in the alphabets of $X_1$
and $Y_1$ amounts to permuting the columns of $A$ and rows of $B$,
respectively. Relabelling of elements in the alphabet of $U$, and hence
the corresponding elements in the alphabet of $V$, requires
simultaneous permutation of the rows of $A$ and columns of $B$. It is
obvious that all these relabellings, and hence the corresponding
permutations of rows and columns of $A$ and $B$, do not change the
underlying system model. Therefore we will implicitly assume any such
convenient permutations in the rest of the paper.

As mentioned before, detection of maliciousness of the relay is to be
done during normal data transmission. That is, the detection is based
upon the symbols that node 1 transmits at time instants $1,2,\ldots,N$
(i.e., $x_1^N$) and the corresponding symbols that node 1 receives at
time instant $N+1,N+2,\ldots,2N$ (i.e., $y_1^N$). We refer to each
pair of such corresponding transmit and receive symbols (e.g., the
ones at time instants $1$ and $N+1$) as a single \emph{observation}
made by node 1. Node 1 is free however to use its $N$ observations to
detect maliciousness of the relay. For instance, we may employ the
following estimator of $\Phi^N$ to construct a decision statistic for
maliciousness detection. First node 1 obtains the conditional
histogram estimator $\hat\Gamma^N$ of $\Gamma^N$ defined by its
$(i,j)$th element:
\begin{equation}\label{e:histo}
\hat\Gamma^N_{i,j} \triangleq \frac{\pi^N(y_{1,i},
  x_{1,j})}{\pi^N(x_{1,j})}.
\end{equation}
Then it constructs the estimator $\hat\Phi^N$ from $\hat\Gamma^N$
according to:
\begin{equation}\label{e:maxhatPhi}
  \hat\Phi^{N} = 
  \begin{cases}
    \displaystyle \arg\hspace*{-12pt}\max_{\hat\Phi \in
      G_{\mu}(\hat\Gamma^{N})} \hspace*{-6pt}\labs{\hat\Phi - I} &
    \mbox{if~}
    G_{\mu}(\hat\Gamma^N) \mbox{~is non-empty,}\\
    I & \mbox{otherwise.}
  \end{cases}
\end{equation}
In (\ref{e:maxhatPhi}), $\mu$ is a small positive constant and
$G_{\mu}(\hat\Gamma)$ is the set of $\abs{U}\!\times\!\abs{U}$
stochastic matrices, for each of which (say denoted by $\hat\Phi$),
there exists a $\abs{Y_1}\!\times\!\abs{U}$ stochastic matrix
$\tilde\Gamma$ such that $\labs{\Pi_B (\tilde\Gamma -
  \hat\Gamma)\Pi_A} \leq \mu$ and $B \hat\Phi A = \Pi_B \tilde\Gamma
\Pi_A$, where $\Pi_A$ and $\Pi_B$ denote the orthogonal projectors
onto the row space of $A$ and column space of $B$, respectively.  We
will employ the estimator $\hat\Phi^N$ specified in
(\ref{e:maxhatPhi}) to obtain the detectability results in the
following sections.

\section{Maliciousness Detectability} \label{sec:main}

To describe the main results of this paper, we first need to introduce
the following notions of normalized, balanced, and polarized vectors:
\begin{define} \textbf{(Normalized vector)}
A non-zero vector $\omega$ is said to be normalized if
  $\labs{\omega} = 1$.
\end{define}
\begin{define} \textbf{(Balanced vector)}
A vector $\omega$ is said to be balanced if
$\sum_i \omega_i = 0$.
\end{define}
\begin{define} \textbf{(Polarized vectors)} \label{defn:pv} For $b
  \geq 0$ and $0\leq \epsilon \leq b$, a vector $\omega$ is said to be
  $(b,\epsilon)$-polarized at $j$ if
\[
  \omega_{i} \in
\begin{cases}
  [b,\infty) & \mbox{~for~}i=j \\
  (-\infty,\epsilon] & \mbox{~for~} i \neq j.
\end{cases}
\] 
Further $\omega$ is said to be $(b,\epsilon)$-double polarized at
$(j,k)$ if 
\[
\omega_{i} \in
\begin{cases}
  [b,\infty) & \mbox{~for~}i=j \\
  (-\infty,-b] & \mbox{~for~}i=k \\
  [-\epsilon,\epsilon] & \mbox{~for~} i \neq j,k.
\end{cases}
\]
\end{define}

\subsection{Main result}
The main result of this paper is that detectability of maliciousness
of the relay is characterized by the following categorization of
observation channels:
\begin{define} \label{def:manipulable} \textbf{(Manipulable
    observation channel)} The observation channel $(A,B)$ is
  manipulable if there exists a $\abs{U}\!\times\!\abs{U}$ non-zero
  matrix $\Upsilon$, whose $j$th column, for each
  $j=1,2,\ldots,\abs{U}$, is balanced and $(0,0)$-polarized at $j$,
  with the property that all columns of $\Upsilon A$ are in the right
  null space of $B$.  Otherwise, $(A,B)$ is said to be
  non-manipulable.
\end{define}
Let $D^{N} = D^{N}(y_1^N,x_1^N)$ denote a decision statistic based on
the first $N$ observations $(y_1^N,x_1^N)$ that is employed for
maliciousness detection. The following theorem states that
maliciousness detectability is equivalent to non-manipulablility of
the observation channel: 
\begin{theorem}\label{thm:main}\textbf{(Maliciousness detectability)}
  When and only when the observation channel $(A,B)$ is
  non-manipulable, there exists a sequence of decision statistics
  $\{D^N\}$ with the following properties (assuming $\delta>0$ below):
\begin{enumerate}
\item If $\limsup_{N\rightarrow\infty} \Pr(\labs{\Phi^{N} - I} >
  \delta) >0$, then
\[
\limsup_{N\rightarrow\infty} \Pr \left( D^{N} > \delta ~\Big|~
  \labs{\Phi^{N} - I} > \delta \right) = 1.
\]
\item If $\liminf_{N\rightarrow\infty} \Pr(\labs{\Phi^{N} - I} \leq
  \delta) >0$, then
\[
\lim_{N\rightarrow\infty} \Pr\left( D^{N} > c\delta~\Big|~
  \labs{\Phi^{N} - I} \leq \delta \right) = 0
\]
for some positive constant $c$ that depends only on $A$ and $B$.
\end{enumerate}
\end{theorem}
The theorem verifies the previous claim that the attack channel
$\Phi^N$ provides us the required statistical characterization for
distinguishing between malicious and non-malicious amplify-and-forward
relay.  In addition, as shown in the proof of the theorem to be
provided later in Section~\ref{sec:proofmain} this distinguishability
is (asymptotically) observable through the decision statistic
$\labs{\hat\Phi^N-I}$, where $\hat\Phi^N$ is the estimator of $\Phi^N$
described in (\ref{e:maxhatPhi}), for non-manipulable channels.  The
requirement of $(A,B)$ being non-manipulable is not over-restrictive
and is satisfied in many practical scenarios.

Theorem~\ref{thm:main} can further be employed to characterize
detectability of maliciousness of the relay in the context of
Definition~\ref{def:maliciousness}:
\begin{cor} \label{thm:maincor} Given that $(A,B)$ is non-manipulable,
  the sequence of decision statistics $\{D^N\}$ in
  Theorem~\ref{thm:main} also satisfies the following properties:
\begin{enumerate}
\item If the relay is not malicious (i.e., $\labs{\Phi^N - I}
  \rightarrow 0$ in probability), then
\[
\lim_{N \rightarrow \infty} \Pr ( D^{N} > \delta) = 0
\]
for any $\delta>0$.
\item If the relay is malicious (i.e., $\labs{\Phi^N - I}$ does not
  converge to $0$ in probability), then
\[
  \limsup_{N\rightarrow\infty} \Pr(D^{N} > \delta) \geq
  \limsup_{N\rightarrow\infty} \Pr(\labs{\Phi^{N} - I} > \delta)
\]
for any $\delta >0$.
\item If the relay is malicious and there is a subsequence
  $\{\Phi^{N_M}\}$ of attack channels satisfying $\labs{\Phi^{N_M} -
    \Phi} \rightarrow 0$ in probability for some stochastic $\Phi \neq
  I$, then there exists $\delta>0$ such that
  $\limsup_{N\rightarrow\infty} \Pr(\labs{\Phi^{N} - I} > \delta) =
  1$, and for every such $\delta$,
\[
\limsup_{N \rightarrow \infty} \Pr ( D^{N} > \delta) = 1.
\]

\item If the relay is malicious with $\labs{\Phi^N - \Phi} \rightarrow
  0$ in probability for some stochastic $\Phi \neq I$, then there
  exists $\delta>0$ such that $\lim_{N\rightarrow\infty}
  \Pr(\labs{\Phi^{N} - I} > \delta) = 1$, and for every such $\delta$,
\[
\lim_{N \rightarrow \infty} \Pr ( D^{N} > \delta) = 1.
\]
\end{enumerate}
\end{cor}
Note that Properties~1 and 2 of the corollary together state that $D^N
\rightarrow 0$ in probability when and only when the relay in not
malicious. Properties~3 and 4 provide progressively stronger
maliciousness detection differentiation when more restrictions are
placed on the attack channel $\Phi^N$.

\subsection{Checking for non-manipulability}
\label{sec:check}

The manipulability of the observation channel $(A,B)$ can be checked
by solving a linear program as shown below:
\begin{alg} \label{thm:check_mani} \textbf{(Non-manipulable?)}
\begin{enumerate}
\item Let $\Omega$, $\nu$, and $\lambda$ be a
  $\abs{X_1}\!\times\!\abs{Y_1}$ matrix-valued variable and two
  $1\!\times\!\abs{U}$ vector-valued variables, respectively.
\item Solve the following linear program: \label{step:opt}
\begin{align*}
  &\min_{\lambda,\nu,\Omega} ~
  \sum_{k=1}^{\abs{U}} 
  \lambda_{k} - \nu_{k} - [A \Omega B]_{k,k} &\\
  & \mathrm{\; subject\ to} & \\
  &\quad 1- \lambda_k \leq 0 & k=1,2,\ldots,\abs{U}, \\
  & \quad  \nu_k + [A\Omega B]_{k,k}- \lambda_k \leq 0 &
  k=1,2,\ldots,\abs{U},  \\
  & \quad  \nu_k + [A\Omega B]_{k,l} \leq 0 & k\neq l=1,2,\ldots,\abs{U}.
\end{align*}
\item If the optimal value in \ref{step:opt}) is $0$, then conclude
  that $(A,B)$ is non-manipulable. Otherwise (i.e., the optimal value
  is positive), conclude that $(A,B)$ is manipulable.
\end{enumerate}
\end{alg}

For cases where the right null space of $B$ is trivial, checking
manipulability of $(A,B)$ is made simple by Theorem~\ref{thm:B=I}
below.  For notation clarity in expressing the theorem, let us define
the following constants that depend only on $A$:
\begin{align*}
  A_{\min} &\defn  \min_{i} \sum_{j} A_{i,j} \\
  a_{\min} &\defn \frac{A_{\min}}{\abs{U} \left(\abs{X_1} +A_{\min}
    \right)}.
\end{align*}
Note that both $A_{\min}$ and $a_{\min}$ are positive since $A$ does
not contain any all-zero row.
\begin{theorem} \label{thm:B=I}
  Suppose that the right null space of $B$ is trivial. Then $(A,B)$ is
  non-manipulable if and only if the left null space of $A$ does not
  contain any normalized, $(a_{\min},0)$-double polarized vectors.
\end{theorem}
We remark that the condition of non-existence of normalized,
$(a_{\min},0)$-double polarized vectors in the left null space of $A$
is relatively easy to check by for instance employing the following
algorithm:
\begin{alg} \label{thm:Aalg} \textbf{(Double polarized vector in left
    null space?)} Let $n = \abs{U} - \textrm{rank}(A)$. The following
  steps can be employed to check whether the left null space of $A$
  contains any normalized, $(a_{\min},0)$-double polarized vectors:
\begin{enumerate}
\item If $n = 0$, then the left null space of $A$ must not contain any
  normalized, $(a_{\min},0)$-double polarized vector.
\item If $n=\abs{U}-1$, then the left null space of $A$ must contain a
  normalized, $(a_{\min},0)$-double polarized
  vector. \label{step:n=U-1}
\item If $1 \leq n \leq \abs{U}-2$:
\begin{enumerate}
\item Find a $n\!\times\!\abs{U}$ matrix $\Upsilon$ whose rows form a
  basis for the left null space of $A$.
\item Perform elementary row operations, permuting columns if
  necessary, to make $\Upsilon$ into the row-reduced echelon form
  $\Upsilon = ( I~\tilde\Upsilon )$, where $\tilde\Upsilon$ is a
  $n\!\times\!(\abs{U}-n)$ block.
\item For each $i=1,2,\ldots,n$, if all elements of
  $\tilde\Upsilon_i$, except for a single negative element, are zero,
  then go to \ref{step:dpv}). \label{step:allzero}
\item For each $i,j \in\{1,2,\ldots,n\}$ and $i\neq j$, if
  $\tilde\Upsilon_i = c \tilde\Upsilon_j$ for some $c>0$, then go to
  \ref{step:dpv}). \label{step:=}
\item Conclude that the left null space of $A$ does not contain any
  normalized, $(a_{\min},0)$-double polarized vector, and terminate.
\item Conclude that the left null space of $A$ contains a normalized,
  $(a_{\min},0)$-double polarized vector. \label{step:dpv}
\end{enumerate}
\end{enumerate}
\end{alg}
Practically speaking, the triviality of the right null space of $B$
guarantees that the pmf of $V$ can be unambiguously obtained by node 1
from observing $Y_1$. This requirement is reasonable if node 1 is
expected to be able to observe the behavior of the relay, and is often
satisfied in practical scenarios. The requirement of the left null
space of $A$ not containing any normalized double-polarized vector is
not over-restrictive, and can be satisfied in many cases by adjusting
the source distribution of node~2.

\section{Numerical Examples} \label{sec:examples}

\subsection{Motivating example}
To illustrate the use of Theorem~\ref{thm:main}, let us first
reconsider the motivating example in Section~\ref{se:motivate}. In the
notation of Section~\ref{sec:model}, $X_1$ and $X_2$ have the same
binary alphabet $\{0,1\}$, and the MAC is the binary erasure MAC
described by $U=X_1+X_2$. That is, the alphabets of $U$ and $V$ are
both $\{0, 1, 2\}$. The BC is ideal defined by $Y_1=V$ and $Y_2=V$. In
addition, we assume the usual equally likely source distributions,
i.e., $X_{1}$ and $X_{2}$ are i.i.d. equally likely binary random
variables.  Physically, this model approximates the scenario in which
two equal-distance Ethernet nodes send signals (collision) to a bridge
node, or the scenario in which two power-controlled wireless nodes
send phase synchronized signals (collision) to an access point. In
both scenarios, the signal-to-noise ratio is assumed to be high.

It is easy to check that in this case
\[
A = \left( \begin{matrix}
.5 &0 \\
.5 & .5 \\
0& .5
\end{matrix} \right) \text{~~and~~} B=I_{3\!\times\! 3}.
\]
Hence $A_{\min}=\frac{1}{2}$, $a_{\min}=\frac{1}{15}$, and
$b_{\min}=\frac{1}{12}$. Note that the left null space of $A$ has
dimension $1$, and the row-reduced echelon basis matrix in
Algorithm~\ref{thm:Aalg} is $(1\ -1\ 1)$. Thus
Algorithm~\ref{thm:Aalg} gives the fact that the left null space of
$A$ does not contain any normalized, $(a_{\min},0)$-double polarized
vector.  By Theorem~\ref{thm:main} and its proof in
Section~\ref{sec:proofs}, we know that the sequence of decision
statistics $\{\labs{\hat\Phi^N - I}\}$, where $\hat\Phi^N$ is
described in (\ref{e:maxhatPhi}), satisfies properties 1) and 2)
stated in Theorem~\ref{thm:main}.  Thus any malicious relay
manipulation is detectable asymptotically.


\begin{figure*} 
  \centering
  \hspace*{-30pt}
  \subfloat[i.i.d attacks, $N=10^{3}$]{\label{n1000_max}
    \includegraphics[width=0.57\textwidth]{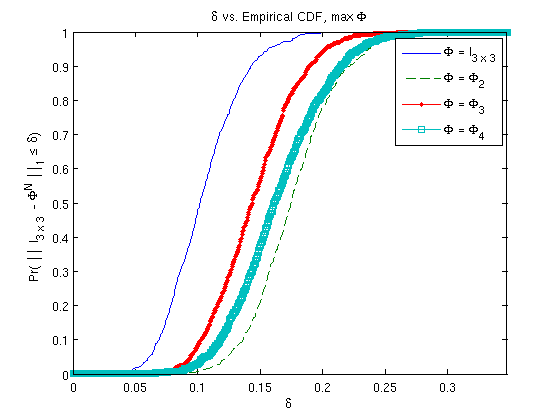}} 
  \hspace*{-30pt}
  \subfloat[i.i.d attacks, $N=10^4$]{\label{n10000_max}
    \includegraphics[width=0.57\textwidth]{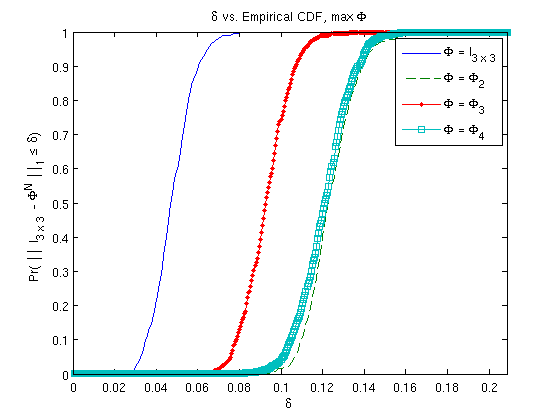}} 

  \hspace*{-30pt}
  \subfloat[i.i.d attacks, $N=10^5$]{\label{n100000_max}
    \includegraphics[width=0.57\textwidth]{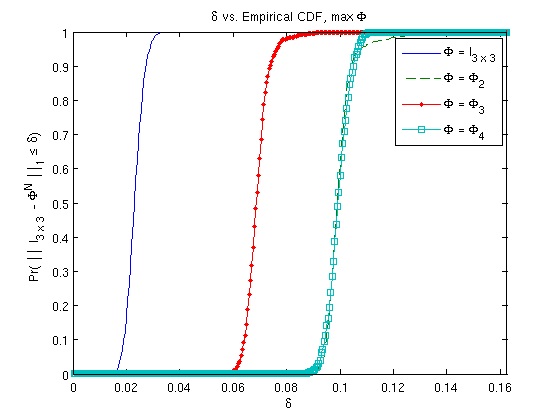}}
  \hspace*{-30pt}
  \subfloat[non-ergodic attacks, $N=10^5$]{\label{b3x3_ne}
    \includegraphics[width=0.57\textwidth]{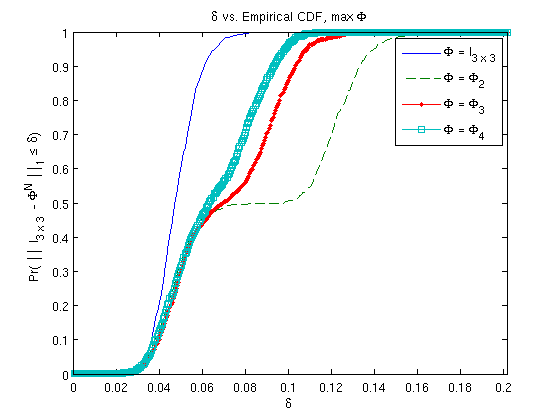}}
  \caption{Plot of empirical cdfs of $\labs{ \hat\Phi^N - I}$
  obtained in the motivating example with the four relay manipulation
  maps corresponding to $\Phi_1=I, \Phi_2$, $\Phi_3$, and $\Phi_4$
  respectively.}
\label{fig:cdf}
\end{figure*}

\subsubsection{I.i.d. attacks} \label{sec:iid}
To demonstrate the asymptotic maliciousness detection performance
promised by Theorem~\ref{thm:main}, and to investigate the performance
with finite observations, we performed simulations for four different
manipulation maps which correspond to the relay randomly and
independently switching its input symbol by symbol according to the
conditional pmfs $p(v_j|u_i)$ specified by the matrices
\begin{align}
&\Phi_1=I_{3\!\times\! 3}, & \Phi_2 = \left(
\begin{matrix}
 .99 & . 005 & .005 \\
.005 & .99 & .005 \\
.005 & .005 & .99
\end{matrix}
\right), \nonumber \\
&\Phi_3  = \left(
\begin{matrix}
  .99 & .005  & 0 \\
  .01 & .99 & .01 \\
  0 & .005 &.99\\
\end{matrix}
\right),
&\text{and~~}
\Phi_4  = \left(
\begin{matrix}
  .99 & 0  & 0 \\
  .01 & 1 & .01 \\
  0 & 0 &.99\\
\end{matrix}
\right).
\label{e:motivPhi}
\end{align}

Obviously the case of $\Phi_1$ corresponds to a non-malicious relay
and the other cases correspond to a malicious relay. The particular
malicious $\Phi$'s were chosen to represent different ways a relay
node may choose to attack. For the case of $\Phi_{2}$, the relay node
changes $1$\% of the symbols received without regards to whether or
not this will make the manipulation obvious to the source nodes.  In
contrast, the attack of $\Phi_{3}$ is more reserved in what it will
do.  It can be seen that the relay's manipulation can only be
instantly detected by one of the source nodes at any given transmitted
value. That is, the relay the relay switches $1$\% of the received
symbols in ways listed out in Table.~\ref{tb:motivate_outcome} except
those labeled with ``Alice \& Bob both detect.''  Finally the attack
of $\Phi_{4}$ is the most cautious and will only take an action that
neither source node can recognize as manipulation without looking at
multiple observations. For this case, the relay switches $2$\% of the
received symbols with values $0$ or $2$ to $1$. This corresponds to
the ``Not detected'' outcomes in Table~\ref{tb:motivate_outcome}.
Note that $\labs{\Phi^N - \Phi_i} \rightarrow 0$ in probability as $N
\rightarrow \infty$ in each case. Hence property 1) of
Corollary~\ref{thm:maincor} applies for the case of $\Phi_1$ and
property 4) applies for the cases of $\Phi_2$, $\Phi_3$, and $\Phi_4$.

In different simulation runs, we set $N = 10^{3}$, $10^{4}$, and
$10^{5}$.  Five thousand trials were run in each simulation. The
empirical cumulative distribution functions (cdfs) of $\labs{
  \hat\Phi^{N} - I}$ obtained from the $5000$ trials for each
simulation are plotted in Figs.~\ref{n1000_max}, \ref{n10000_max}, and
\ref{n100000_max} for the cases of $N=10^{3}$, $10^{4}$, and $10^{5}$,
respectively.  For these three cases, the values of $\mu$ chosen in
defining the estimator $\hat\Phi^N$ are $0.2$, $0.1$, and $0.05$,
respectively. From Figs.~\ref{n10000_max} and \ref{n100000_max}
respectively with $N=10^4$ and $N=10^5$, as predicted by parts 1) and
4) of Corollary~\ref{thm:maincor}, the decision statistic $\labs{
  \hat\Phi^{N} - I}$ succeeds in differentiating between the
non-malicious case of $\Phi_1$ and the malicious cases of $\Phi_2$,
$\Phi_3$, and $\Phi_4$ with very high confidence. For instance, by
selecting the decision threshold at $\delta = 0.065$ and $\delta =
0.004$ respectively for the cases of $N=10^4$ and $N=10^5$, we are
able to obtain very small miss and false alarm probabilities for
detecting maliciousness of the relay.  For $N=10^3$, we can see from
Fig. \ref{n1000_max} that there is still differentiation between the
empirical cdfs obtained for the non-malicious and malicious
cases. However the maliciousness differentiation confidence achieved
is much weaker than the detectors with the larger value of $N$. This
simulation exercise illustrates the fact that the decision statistic
based on the maximum-norm estimator of (\ref{e:maxhatPhi}), while is
convenient for proving the asymptotic distinguishability result in
Theorem~\ref{thm:main}, may not be a suitable choice for constructing
a practical detector when the number of observations, $N$, is not
large. Other more efficient finite-observation detectors may be
needed.

\subsubsection{Non-ergodic attacks}
To demonstrate part 2) of Corollary~\ref{thm:maincor} with non-i.i.d
attacks, we simulated a few non-ergodic attacks and considered again
the decision statistic $\labs{ \hat\Phi^{N} - I}$.  In these
non-ergodic attacks, the relay decides whether or not to manipulate
the symbols depending on if the checksum of all observed symbols is
even or not. Conditioning on an even checksum, the relay manipulates
the symbols i.i.d. according to $\Phi_2$, $\Phi_3$, and $\Phi_4$ as
described in (\ref{e:motivPhi}).  Note that the for these attacks,
$\lim_{N\rightarrow \infty} \Pr (\labs{\Phi^N - I} > \delta) = 0.5$
for all $\delta>0$.

The results for this simulation with $N=10^5$ and $\mu = 0.01$ are
plotted in Fig.~\ref{b3x3_ne}. From the figure, the first important
note is that the empirical cdfs of the decision statistic exhibit
staircase shapes with a step at $0.5$ as predicted by part 2) of
Corollary~\ref{thm:maincor}.  When the relay is being malicious, it is
clear that by choosing $\delta = 0.07$, we can again obtain small miss
and false alarm probabilities for detecting maliciousness of the
relay.

\subsection{Higher order example} \label{sec:highorder}
\begin{figure*}
\hrulefill
\begin{align}
& A  = \left(
\begin{matrix}
  \frac{1}{3} & 0  & 0  \\
  \frac{1}{3} & \frac{1}{3} & 0  \\
  \frac{1}{3} & \frac{1}{3} &\frac{1}{3}\\
  0 & \frac{1}{3} & \frac{1}{3} \\
  0 & 0 & \frac{1}{3} \\
\end{matrix}
\right),~  
B =\left(
\begin{matrix}
  1 & .5 &0 & 0 & 0 \\
  0 & .5 & .7 & 0 & 0 \\
  0 & 0 &.3 & .5 & 0\\
  0 & 0 & 0 & .5 & 1\\
\end{matrix}
\right), ~\Phi_1=I_{5\!\times\! 5},
~
\Phi_2  =
\left(
\begin{matrix}
 .99 & . 0025 & .0025 & .0025 & .0025 \\
.0025 & .99 & .0025 & .0025 & .0025 \\
.0025 & .0025 & .99 & .0025 & .0025\\
.0025 & .0025 & .0025 & .99 & .0025 \\
.0025 & .0025 & .0025 & .0025 & .99 \\
\end{matrix}
\right),  \nonumber \\
&\Phi_3  = \left(
\begin{matrix}
 .99 & \frac{.01}{3} & .0025 & 0 & 0 \\
.005 & .99 & .0025 & \frac{.01}{3} & 0 \\
.005 & \frac{.01}{3} & .99 & \frac{.01}{3} & .005\\
0 & \frac{.01}{3} & .0025 & .99 & .005 \\
0 & 0 & .0025 & \frac{.01}{3} & .99 \\
\end{matrix}
\right), 
\text{~and~}
\Phi_4  = \left(
\begin{matrix}
 .985 & 0 & 0 & 0 & 0 \\
.0075 & .985 & 0 & 0 & 0 \\
.0075 & .015 & 1 & .015 & .0075\\
0 & 0 & 0 & .985 & .0075 \\
0 & 0 & 0 & 0 & .985 \\\end{matrix}
\right).
\label{e:highorder}
\end{align}
\hrulefill
\end{figure*}

\begin{figure*}
  \centering
  \hspace*{-30pt}
  \subfloat[$(A,B)$ non-manipulable]{\label{b4x5_1e6}
    \includegraphics[width=0.57\textwidth]{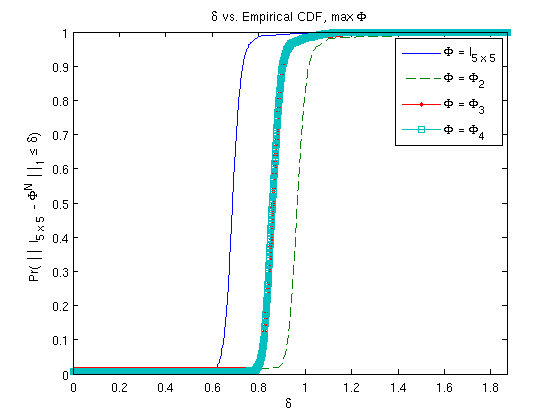}}
  \hspace*{-20pt}
  \subfloat[$(A,B)$ manipulable]{\label{b4x5_counter}
    \includegraphics[width=0.49\textwidth]{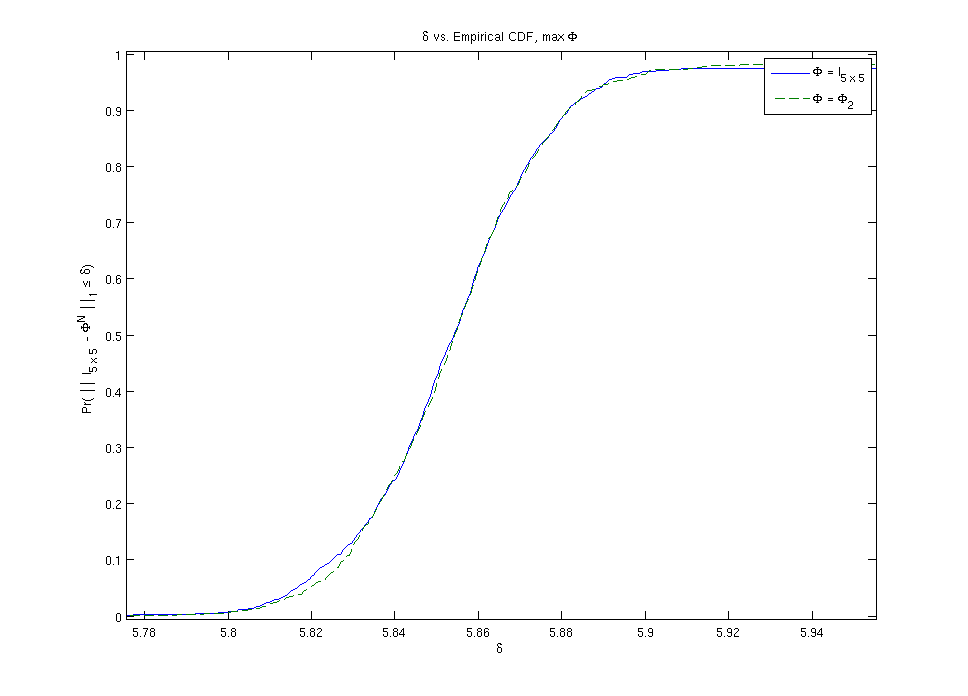}}
  \caption{Plot of empirical cdfs of $\labs{ \hat\Phi^N - I}$ for
    various i.i.d. attacks considered in Sections~\ref{sec:highorder}
    and \ref{sec:counter}.}
\label{fig:B4x5}
\end{figure*}

Let us consider the addition channel as shown in
Fig.~\ref{fig:motivate} with both Alice and Bob choosing their source
symbols uniformly over the ternary alphabet $\{0,1,2\}$ instead. Hence
the input and output alphabets of Romeo is $\{0,1,2,3,4\}$ in this
case. It is easy to verify that the corresponding $A$ matrix is given
as in (\ref{e:highorder}) on the next page.  Furthermore, suppose that
BC from Romeo back to Alice and Bob is not ideal. In particular, let
us model the marginal BC from Romeo back to Alice by the matrix $B$
given in (\ref{e:highorder}). Notice that this $B$ has a non-trivial
right null space.

First we need to determine if the pair $(A,B)$ is non-manipulable. To
do this, we used the Algorithm~\ref{thm:check_mani} presented in
Section~\ref{sec:check}.  In particular, we employed the linear
programming solver \texttt{linprog} in the optimization toolbox in
MATLAB to solve the linear program in step~\ref{step:opt} of
Algorithm~\ref{thm:check_mani}. The optimal value returned was of the
order of $10^{-16}$, which is close enough to $0$ for us to decide
$(A,B)$ as non-manipulable. Thus again Theorem~\ref{thm:main} and
Corollary~\ref{thm:maincor} apply to give that the decision statistic
$\labs{\hat\Phi^N -I}$ provides maliciousness detectability for this
channel.

As in Section~\ref{sec:iid}, we simulated i.i.d. attacks by Romeo. The
four different $\Phi$'s shown in (\ref{e:highorder}) were the cases
that we considered in the simulation study.  The attack of $\Phi_1$
corresponds to the case in which Romeo truthfully forwards the
received symbols. For the attacks of $\Phi_2$, $\Phi_3$, and $\Phi_4$,
Romeo alters $1$\% of the symbols that it receives. Each of these
three cases was once again chosen for a particular level of
maliciousness as in Section~\ref{sec:iid}. The case of $\Phi_2$
corresponds to an attack in which Romeo returns values that he knows
will instantly guarantee detection. The attack of $\Phi_3$ only sends
back values for which it is possible to not be instantly
detected. Finally $\Phi_4$ corresponds to the case in which Romeo is
the most cautious, and will not send back any symbol which is
instantly detectable. 

The empirical cdfs of $\labs{\hat \Phi ^N - I}$ obtained for the four
different $\Phi_i$'s are plotted in Fig.~\ref{b4x5_1e6} for the
simulation run with $N=10^5$ and $\mu=0.05$. As before, by choosing a
decision threshold at $\delta = 0.07$ we can obtain very small miss
and false alarm probabilities for detecting maliciousness of Romeo.

\subsection{Counter-example} \label{sec:counter}

To demonstrate the consequence of having a manipulable observation
channel $(A,B)$, reconsider the ternary-input example of
Section~\ref{sec:highorder} with the marginal BC from Romeo back to
Alice specified by the following matrix
\[
B  =\left(
\begin{matrix}
  1 & 0 & 0 & 0 & 0 \\
  0 & .5 & 0 & .3 & .2 \\
  0 & 0 &.5 & .2 & .3\\
  0 & .3 & .2 & .5 & 0\\
    0 & .2 & .3 & 0 & .5\\
\end{matrix}\right).
\]
To check whether $(A,B)$ is manipulable,
Algorithm~\ref{thm:check_mani} was again employed. The optimal value
of the linear program in step~\ref{step:opt} obtained was $3$, thus
alerting us that $(A,B)$ is manipulable.
Indeed it can be readily check that for any $\psi\in (0,1]$, the
matrix  
\[
\Upsilon  =\left(
\begin{matrix}
  0 & 0 & 0 & 0 & 0 \\
  0 & \psi & 0 & 0 & 0 \\
  0 & 0 & \psi  & 0 & -\psi\\
  0 & -\psi & 0 & 0 & 0\\
    0 & 0 & -\psi & 0 & \psi\\
\end{matrix}\right), 
\]
is one that satisfies $B \Upsilon A = 0$ required in
Definition~\ref{def:manipulable} to make $(A,B)$
manipulable. Therefore Theorem~\ref{thm:main} tells us that
maliciousness detectability is impossible for this channel.

As in Sections~\ref{sec:iid} and \ref{sec:highorder}, an i.i.d. attack
with $\Phi_2 = I - \Upsilon$ was simulated for $N=10^5$ and
$\mu=0.05$. The value of $\psi = 1$ was chosen in the simulation. This
choice corresponds to an average of $\frac{5}{9}$ of the symbols are
changed by Romeo. Clearly having this many symbols changed would be
catastrophic in most practical communication systems, and is therefore
undesirable. The empirical cdfs of the non-malicious case and the
malicious case of $\Phi_2$ obtained from the simulation are plotted in
Fig.~\ref{b4x5_counter}. It is clear from the figure that the cases
for which Romeo is being malicious and not malicious are
indistinguishable. Hence the severe attack of $\Phi_2$ can not be
detected.

\section{Proofs of Detectability Results} \label{sec:proofs} 

In order to prove the various results in Section~\ref{sec:main}, we
will need to extend the notion of polarization of vectors given in
Definition~\ref{defn:pv} to matrices:
\begin{define}\textbf{(Polarized matrices)}
  Let $1 \leq n\leq\abs{U}$.  For $b\geq 0$ and $0\leq \epsilon \leq
  b$, we say that a $n\!\times\!\abs{U}$ matrix $\Upsilon$ is
  $(b,\epsilon)$-polarized if
\[
\Upsilon_{i,j} \in 
\begin{cases}
  [b,\infty) & \mbox{~for~}j=i \\
  (-\infty,\epsilon] & \mbox{~for~} j \neq i.
\end{cases}
\]
If, in addition, $\Upsilon_{i,j} =0$ for all $i,j = 1,2,\ldots,n$ and
$i\neq j$, we say that $\Upsilon$ is $(b,\epsilon)$-diagonal
polarized.  
\end{define}
Moreover, by saying a normalized $\Upsilon$ is in the left null space
of $A$, we mean all rows of $\Upsilon$ are normalized vectors in the
left null space of $A$.

In addition to the notion of polarized vectors and matrices, we will
also employ the following generalization of linear dependence: 
\begin{define} \textbf{($\epsilon_s$-dependent)} 
  A vector $\omega$ is $\epsilon_s$-dependent ($\epsilon_s \geq 0$) upon
  a set of vectors $\Upsilon_1,\Upsilon_2,\ldots,\Upsilon_n$ of the
  same dimension if there exists a set of coefficients
  $c_1,c_2,\ldots,c_n$ such that
\[
\left\|\omega - \sum_{i=1}^{n} c_i \Upsilon_{i}\right\|_1 \leq
\epsilon_s.
\]
\end{define}

With these definitions in place, we will first establish a few
important and interesting properties of polarized vectors and matrices
in the left and right null spaces of $A$ and $B$, respectively. In
addition, we will show that the condition of the observation channel
$(A,B)$ being non-manipulable is sufficient in guaranteeing the
validity of these properties.  Then we will apply some of these
properties to bound the distance between an estimate of the attack
channel and the true attack channel. The distance bound is employed to
show that a decision statistic constructed from an attack channel
estimator based on histogram estimation of node 1's conditional pmf of
its received symbols given its transmitted symbols provides the needed
convergence properties in Theorem~\ref{thm:main}. The aforementioned
properties of polarized vectors in the left null space of $A$ will
also be used to prove Algorithm~\ref{thm:Aalg} and
Theorem~\ref{thm:B=I}.

\subsection{Properties concerning polarized vectors and matrices in
  null spaces of $A$ and $B$}
\label{sec:lemmas}

Let us first study the left null space of $A$.  The following simple
lemma about normalized vectors in the left null space of $A$ is
critical to many other results in this section:
\begin{lemma}\label{thm:lumaxnull} 
  Suppose that $\upsilon$ is a non-zero normalized
  $1\!\times\!\abs{U}$ vector in the left null space of $A$.  Then
  $\upsilon$ must contain at least one positive element and one
  negative element, and
\begin{align*}
\max_{i:\upsilon_i > 0} \upsilon_i &\geq  a_{\min} \\
\min_{i:\upsilon_i < 0} \upsilon_i &\leq -a_{\min}.
\end{align*}
\end{lemma}
\begin{IEEEproof}
  Write $a=\max_{i:\upsilon_i > 0}\upsilon_i$ for convenience, and
  note that $a=0$ by definition if $\upsilon$ contains no positive
  element.  Since $\upsilon$ is normalized, we have
\begin{equation}
  \sum_{i: \upsilon_{i} \leq 0} \abs{\upsilon_{i}}
  = 1- \sum_{i: \upsilon_{i} > 0} \abs{\upsilon_{i}}
  \geq 1-a \abs{U}.
\label{e:negbound}
\end{equation}
Because $\upsilon$ is in the left null space of $A$,
$\sum_{i}\upsilon_{i} A_{i,j} = 0$ for all $j$. That implies
\begin{equation}
  \sum_{i:\upsilon_{i} > 0} \upsilon_{i}A_{i,j} 
  = -\sum_{i:\upsilon_{i} \leq 0} \upsilon_{i}A_{i,j}
  = \sum_{i: \upsilon_{i} \leq 0} \abs{\upsilon_{i}}A_{i,j}.
\label{e:posnegnull}
\end{equation}
But, because $0\leq A_{i,j}\leq 1$, we can make the following
inequality
\[
\sum_{i: \upsilon_{i} > 0} \upsilon_{i}A_{i,j} \leq \sum_{i:
  \upsilon_{i} > 0} \upsilon_{i} \leq a \abs{U}.
\]
Substituting (\ref{e:posnegnull}) back in, we get
\[
a\abs{U} \geq \sum_{i: \upsilon_{i} \leq 0} \abs{\upsilon_{i}}A_{i,j}
\]
which must hold for all $j$. Therefore,
\[
a\abs{U}\abs{X_1} \geq \sum_{j}\sum_{i: \upsilon_{i} \leq 0}
\abs{\upsilon_{i}}A_{i,j} \geq \sum_{i: \upsilon_{i} \leq 0}
\abs{\upsilon_{i}}A_{\min}
\]
which causes a contradiction when $a=0$ since $A_{\min}>0$. Hence, $a$
must not be $0$, and $\upsilon$ must have at least one positive
element.  Further, by (\ref{e:negbound}),
\[
a\abs{U}\abs{X_1} \geq A_{\min} \left( 1-a\abs{U}\right)
\]
which gives the desired lower bound on $a$. The proof of existence of
a negative element and the fact that the minimum negative element must
be no larger than $-a_{\min}$ is similar.
\end{IEEEproof}
An immediate, but important later in proving Theorem~\ref{th:bound},
consequence of the lemma is the following observation:
\begin{lemma} \label{thm:offdiag}
  Let $0<\epsilon \leq a_{\min}$. Suppose that $\omega$ is a
  normalized vector in the left null space of $A$ that is not
  $(a_{\min},\epsilon)$-polarized at $i$. Then there exists a $j\neq
  i$ such that $\omega_j \geq \epsilon$.
\end{lemma}
\begin{IEEEproof}
  Since $\omega$ is normalized, $\omega_k \geq a_{\min}$ for some $k$
  by Lemma~\ref{thm:lumaxnull}. If $k \neq i$, then we have the stated
  conclusion. Now suppose $k=i$. If $\omega_j < \epsilon$ for all $j
  \neq i$, then we obtain the contradictory conclusion that $\omega$
  is $(a_{\min},\epsilon)$-polarized at $i$.
\end{IEEEproof}
Lemma~\ref{thm:lumaxnull} also implies the following two lemmas about
normalized, diagonal polarized matrices in the left null space of $A$:
\begin{lemma} \label{t3} 
  Let $a_{\min} \leq a \leq 1$ and $2\leq n \leq \abs{U}$. Let $0 \leq
  \epsilon_s < a_{\min}$, and $\epsilon$ and $\epsilon_{n-1}$ be two
  positive constants. Define
\begin{equation}
  \epsilon'_n=\frac{\epsilon +\epsilon_{s}}{a_{\min} - \epsilon_{s}}
  +(n-1)\frac{\epsilon + \epsilon_{n-1} + \epsilon_{s} +
    \epsilon_{s}\epsilon_{n-1}}{a(a_{\min} - \epsilon_{s})}.
  \label{e:e'}
\end{equation}
Suppose that $\epsilon_s$, $\epsilon$, and $\epsilon_{n-1}$ are chosen
small enough to satisfy $\epsilon'_n < a_{\min}$.  Furthermore,
suppose that the left null space of $A$ contains a normalized,
$(a,\epsilon_{n-1})$-diagonal polarized, $(n-1)\!\times\!\abs{U}$
matrix and a normalized, $1\!\times\!\abs{U}$ vector that is
$(a,\epsilon)$-polarized at $n$, and is $\epsilon_{s}$-dependent on
the rows of the matrix.  Then the left null space of $A$ also contains
a normalized, $1\!\times\!\abs{U}$ vector that is
$(a_{\min},\epsilon')$-double polarized at $(m,n)$ for all
$\epsilon'_n \leq \epsilon'<a_{\min}$, where $m < n$.
\end{lemma}
\begin{IEEEproof}
  Fix $\epsilon' \in [\epsilon'_n,a_{\min})$.  Let $\Upsilon$ and
  $\upsilon$ be the $(a,\epsilon_{n-1})$-diagonal polarized matrix and
  the $(a,\epsilon)$-polarized vector, respectively, given in the
  statement of the lemma. We will show that one row of $\Upsilon$ must
  be $(a_{\min},\epsilon')$-double polarized at $(m,n)$, where $m <
  n$.
  
  First, since $\upsilon$ is $\epsilon_s$-dependent upon the rows in
  $\Upsilon$, we know that there exists a set of coefficients
  $c_{1},c_{2},\ldots,c_{n-1}$ such that for any
  $j=1,2,\ldots,\abs{U}$,
\begin{equation}
  \upsilon_{j} - \epsilon_{s} \leq \sum_{i=1}^{n-1}c_{i} \Upsilon_{i,j} \leq 
  \upsilon_{j} + \epsilon_{s}.
  \label{e:edepupUp}
\end{equation}
In particular, for $i=1,2,\ldots, n-1$, we have $\upsilon_i-\epsilon_s
\leq c_{i} \Upsilon_{i,i} \leq \upsilon_{i}+\epsilon_{s}$. Using the
facts that $\Upsilon_{i,i} \geq a$ and that $-1 \leq \upsilon_{i} \leq
\epsilon$, we can determine that
\begin{equation}
  \frac{-1-\epsilon_s}{a} \leq c_{i} \leq \frac{\epsilon
    +\epsilon_{s}}{a}.
\label{e:cibound}
\end{equation}

From Lemma~\ref{thm:lumaxnull}, we know that there exists an index $m$
such that $\upsilon_{m} \leq -a_{\min}$. To proceed, we want to show
that $m < n$, which we will do through contradiction. Obviously $m
\neq n$. Suppose $m > n$ and consider (\ref{e:edepupUp}) with $j=m$.
Separating the terms with positive and negative $c_i$'s in the
summation and using the upper bound in (\ref{e:edepupUp}), we have
\begin{align}
  \sum_{i: c_{i} < 0 }\abs{c_{i}} \Upsilon_{i,m} &\geq -\upsilon_{m}
  -\epsilon_{s} +
  \sum_{i: c_{i} > 0 }c_{i} \Upsilon_{i,m} \nonumber \\
  &\geq a_{\min} -\epsilon_{s} -(n-1)\frac{\epsilon
    +\epsilon_{s}}{a}
 \label{e:lb1}
\end{align} 
where the second inequality is obtained by using the upper bound on
$c_i$ in (\ref{e:cibound}).  But because $m>n$, we have
$\Upsilon_{i,m} \leq \epsilon_{n-1}$. Then by using the lower bound on
$c_i$ in (\ref{e:cibound}), we get $\sum_{i: c_{i} < 0 }\abs{c_{i}}
\Upsilon_{i,m} \leq (n-1) \epsilon_{n-1}
\frac{1+\epsilon_s}{a}$. Thus we arrive at the conclusion that
\[
(n-1)\epsilon_{n-1}\frac{1+\epsilon_s}{a} \geq a_{\min} -\epsilon_{s}
-(n-1)\frac{\epsilon +\epsilon_{s}}{a},
\]
which clearly violates the condition $\epsilon'_n< a_{\min}$ in the
statement of the lemma.  Therefore $m < n$. Furthermore we have $c_m
\Upsilon_{m,m} \leq \upsilon_m+\epsilon_s\leq -a_{\min}+\epsilon_s
<0$, which implies $c_m<0$ and $\abs{c_m} \geq a_{\min}-\epsilon_s$.
  
Similar to (\ref{e:lb1}), we have, for $j>n$,
\begin{align}
  \sum_{i: c_{i} < 0 }\abs{c_{i}} \Upsilon_{i,j} &\geq -\upsilon_{j}
  -\epsilon_{s} + \sum_{i: c_{i} > 0 }c_{i} \Upsilon_{i,j} \nonumber \\
  &\geq -\epsilon -\epsilon_{s} -(n-1)\frac{\epsilon
    +\epsilon_{s}}{a}
\label{e:lb2}
\end{align}
where the second inequality is due to the fact that $\upsilon_j \leq
\epsilon$.  Further separating the terms with positive and negative
$\Upsilon_{i,j}$ in the sum on the left side of (\ref{e:lb2}), for
$j>n$, we have
\begin{eqnarray}
  \lefteqn{\sum_{i: c_{i} < 0, \Upsilon_{i,j}<0 }\abs{c_{i}}
    \abs{\Upsilon_{i,j}}}\nonumber \\ 
  &\leq& 
  \epsilon +\epsilon_{s} +(n-1)\frac{\epsilon +\epsilon_{s}}{a}
  +\sum_{i: c_{i} < 0, \Upsilon_{i,j}>0 }\abs{c_{i}} \Upsilon_{i,j}
  \nonumber \\
  &\leq& \epsilon +\epsilon_{s} +(n-1) 
  \frac{\epsilon + \epsilon_{n-1} + \epsilon_{s} + \epsilon_{s}\epsilon_{n-1}}{a}
\label{e:lb3}
\end{eqnarray}
where the second inequality results from the bound $\sum_{i: c_{i} <
  0, \Upsilon_{i,j}>0 }\abs{c_{i}} \Upsilon_{i,j} \leq
(n-1)\epsilon_{n-1}\frac{1+\epsilon_s}{a}$, as shown above.  Because
$c_{m} < 0$, we know from (\ref{e:lb3}) that if $\Upsilon_{m,j} < 0$,
then $\abs{c_{m}} \abs{\Upsilon_{m,j}} \leq \epsilon +\epsilon_{s}
+(n-1)\frac{\epsilon + \epsilon_{n-1} + \epsilon_{s} +
  \epsilon_{s}\epsilon_{n-1}}{a}$. Further, by the above derived
result that $\abs{c_{m}} \geq a_{\min} - \epsilon_{s}$, we get
\[
\abs{\Upsilon_{m,j}} \leq \frac{\epsilon +\epsilon_{s}}{a_{\min} -
  \epsilon_{s}} +(n-1)\frac{\epsilon + \epsilon_{n-1} + \epsilon_{s} +
  \epsilon_{s}\epsilon_{n-1}}{a(a_{\min} - \epsilon_{s})} =
\epsilon'_n,
\]
if $\Upsilon_{m,j}<0$.  Therefore, $\abs{\Upsilon_{m,j}} \leq
\epsilon'$ for all $j>n$.

Combining all above results, we observe that $\Upsilon_{m,m} \geq a$,
$\abs{\Upsilon_{m,j}} \leq \epsilon' < a_{\min}$ for $j>n$, and
$\Upsilon_{m,j}=0$ for all other index values of $j$ except $n$. From
Lemma~\ref{thm:lumaxnull}, we must have $\Upsilon_{m,n} \leq
-a_{\min}$. Thus $\Upsilon_{m}$ is $(a_{\min},\epsilon')$-double
polarized at $(m,n)$.
\end{IEEEproof}

\begin{lemma} \label{t2b} 
  Let $a_{\min} \leq a \leq 1$ and $2\leq n \leq \abs{U}$.  Let $0 <
  \epsilon_s < a_{\min}$, and $\epsilon$ and $\epsilon_{n-1}$ be two
  positive constants. Define
  \begin{equation}
  \epsilon''_n=\frac{\epsilon_{n-1}}{a} + \frac{1}{a_{\min}a
    \epsilon_s} \left[ \epsilon + (n-1)\frac{\epsilon +
      \epsilon_{n-1}}{a} \right].
  \label{e:e''}
  \end{equation}
  Suppose that $\epsilon$ and $\epsilon_{n-1}$ are chosen small enough
  to satisfy $\epsilon''_n < a_{\min}$.  Furthermore, suppose that the
  left null space of $A$ contains a normalized,
  $(a,\epsilon_{n-1})$-diagonal polarized, $(n-1)\!\times\!\abs{U}$
  matrix and a normalized, $1\!\times\!\abs{U}$ vector that is
  $(a,\epsilon)$-polarized at $n$, and is not $\epsilon_{s}$-dependent
  on the rows of the matrix.  Then the left null space of $A$ contains
  a normalized, $(a_{\min},\epsilon'')$-diagonal polarized,
  $n\!\times\!\abs{U}$ matrix, for all $\epsilon''_n\leq \epsilon'' <
  a_{\min}$.
\end{lemma}
\begin{IEEEproof}
  Fix $\epsilon'' \in [\epsilon''_n,a_{\min})$.  Let $\Upsilon$ and
  $\upsilon$ be the $(a,\epsilon_{n-1})$-diagonal polarized matrix and
  the $(a,\epsilon)$-polarized vector, respectively, given in the
  statement of the lemma.  We will first construct a normalized,
  $1\!\times\!\abs{U}$ vector $\hat\upsilon$ that is
  $(a_{\min},\epsilon'')$-polarized at $n$ with the additional
  property that $\hat\upsilon_{j} = 0$ for all $j=1,2,\ldots,n-1$.
  Then we apply elementary row operations and normalization to the
  rows of $[\Upsilon^{T} \hat\upsilon^{T}]^{T}$ to obtain the desired
  normalized, $n\!\times\!\abs{U}$, $(a_{\min},\epsilon'')$-diagonal
  polarized matrix.

  First, set 
  \[
  \tilde\upsilon = \upsilon - \sum_{i=1}^{n-1}
  \frac{\upsilon_{i}}{\Upsilon_{i,i}} \Upsilon_{i}.
  \]
  Because $\upsilon$ is not $\epsilon_s$-dependent upon the rows of
  $\Upsilon$, we must have $\labs{\tilde\upsilon} > \epsilon_s$.
  Hence we can normalize $\tilde\upsilon$ to obtain $\hat\upsilon$,
  i.e., $\hat\upsilon = \tilde\upsilon / \labs{\tilde\upsilon}$.
  Clearly, for $j<n$, $\tilde\upsilon_{j} =\upsilon_{j} -
  \frac{\upsilon_{j}}{\Upsilon_{j,j}}\Upsilon_{j,j} = 0$, which
  implies $\hat\upsilon_j=0$.  For $j>n$, write
  \begin{equation}
    \tilde\upsilon_{j} = \upsilon_{j} -
    \sum_{i:\upsilon_{i}>0}\frac{\upsilon_{i}}{\Upsilon_{i,i}}\Upsilon_{i,j}
    +
    \sum_{i:\upsilon_{i}<0}\frac{\abs{\upsilon_{i}}}{\Upsilon_{i,i}}\Upsilon_{i,j}.
   \label{e:tup}
  \end{equation}
  But since $\Upsilon_{i,i} \geq a$ and $\upsilon_{j} \leq \epsilon$
  for $j>n$, we have
  \begin{align*}
    \sum_{i:\upsilon_{i}>0} \frac{\upsilon_{i}}{\Upsilon_{i,i}}
    \Upsilon_{i,j} &\geq \sum_{i:\upsilon_{i}>0}\frac{\epsilon}{a}(-1)
    \geq -(n-1)\frac{\epsilon}{a} \\
    \sum_{i:\upsilon_{i}<0} \frac{\abs{\upsilon_{i}}}{\Upsilon_{i,i}}
    \Upsilon_{i,j} &\leq \sum_{i:\upsilon_{i}<0} \frac{1}{a}
    \epsilon_{n-1} \leq (n-1)\frac{\epsilon_{n-1}}{a}.
  \end{align*}
  Applying these two bounds to (\ref{e:tup}), we get, for $j>n$,
  \[
  \tilde\upsilon_{j}\leq\epsilon +(n-1) \frac{\epsilon + \epsilon_{n-1}}{a},
  \] 
  which implies
  \[
  \hat\upsilon_{j} \leq \frac{1}{\epsilon_{s}} \left[ \epsilon
    +(n-1) \frac{\epsilon + \epsilon_{n-1}}{a} \right] \defn
  \varepsilon \leq \epsilon''_n.
  \]
  Further, note that $\hat\upsilon$ is a normalized vector in the left
  null space of $A$. Hence by Lemma~\ref{thm:lumaxnull} and the fact
  that $\epsilon''_n < a_{\min}$, $\hat\upsilon_{n} \geq a_{\min}$.
  Therefore $\hat\upsilon$ is $(a_{\min},\varepsilon)$-polarized at
  $n$ with the additional property that $\hat\upsilon_{j}= 0$ for
  $j<n$ as claimed.
  
  Next, for $i=1,2,\ldots, n-1$, set
  \[ 
  \tilde\Upsilon_{i} = \Upsilon_{i} -
  \frac{\Upsilon_{i,n}}{\hat\upsilon_{n}} \hat\upsilon.
  \]
  Clearly, $\tilde\Upsilon_{i,n} = 0$ and $\tilde\Upsilon_{i,j} =
  \Upsilon_{i,j}$ for $j<n$ by design.  Thus
  $\labs{\tilde\Upsilon_{i}} \geq a$.  Again, we can normalize
  $\tilde\Upsilon_{i}$ to get $\hat\Upsilon_{i} = \tilde\Upsilon_{i} /
  \labs{\tilde\Upsilon_{i}}$.  Now, consider $j>n$,
  \[ 
  \tilde\Upsilon_{i,j} = \Upsilon_{i,j} -
  \frac{\Upsilon_{i,n}}{\hat\upsilon_{n}}\hat\upsilon_{j} \leq
  \epsilon_{n-1}+ \frac{\max\{\epsilon_{n-1},\varepsilon\}}{a_{\min}}
  = \epsilon_{n-1}+ \frac{\varepsilon}{a_{\min}}
  \]
  where the second inequality is due to $-1\leq \Upsilon_{i,n} \leq
  \epsilon_{n-1}$, $-1\leq \hat\upsilon_{j} \leq \varepsilon$, and
  $\hat\upsilon_n \geq a_{\min}$, and the last equality results from
  the fact that $\abs{U}\epsilon_s<1$. Hence
  \[
  \hat\Upsilon_{i,j} \leq \frac{\epsilon_{n-1}}{a}+
  \frac{\varepsilon}{a_{\min}a} = \epsilon''_n.
  \]
  As discussed before, $\hat\Upsilon_{i,j} = 0$ for all
  $i=1,2,\ldots,n-1$ and $j \neq i \leq n$. Hence, again using
  Lemma~\ref{thm:lumaxnull}, we must have $\hat\Upsilon_{i,i} \geq
  a_{\min}$ since $\epsilon'' < a_{\min}$.  Finally, set
  $\hat\Upsilon_{n} = \hat\upsilon$. The matrix $\hat\Upsilon$
  composed by using
  $\hat\Upsilon_1,\hat\Upsilon_2,\ldots,\hat\Upsilon_{n}$ as rows is
  the desired $(a_{\min},\epsilon'')$-diagonal polarized matrix.
\end{IEEEproof}
To apply Lemmas~\ref{t3} and \ref{t2b}, we need to the following lemma
to select $\epsilon_n$:
\begin{lemma}\label{thm:recur}
  Let $\epsilon_s$ and $\epsilon$ be two positive constants.  Suppose
  that $2 \leq n \leq \abs{U}$.  Define $\tilde a_{n-1} \defn
  a_{\min}^2 \left( 1 + \frac{n-1}{a_{\min}} \right)^{-1}$.  Set
  $\epsilon_i = \epsilon \left( \frac{2}{\tilde a_{n-1}
      \epsilon_s}\right)^{i-1}$ for $i=1,2,\ldots,n$.  If
  \begin{align}
    \epsilon_s & < a_{\min}^2 \tilde a_{n-1} \Bigg\{ a_{\min}^2
    \left[ 1 + \left(\frac{a_{\min} \tilde a_{n-1}}{2}
      \right)^{n-1}\right]
    \nonumber \\
    & ~~~~~~~~~~ + (n-1)(1+a_{\min})\frac{\tilde a_{n-1}^2}{2} +
    a_{\min}\tilde a_{n-1} \Bigg\}^{-1}
  \label{e:es}
\end{align}
and
  $\epsilon < a_{\min} \left(\frac{\tilde a_{n-1} \epsilon_s}{2}
  \right)^{n-1}$, then
  \begin{enumerate}
  \item $\epsilon_1 \leq \epsilon_2 \leq \cdots \leq \epsilon_n <
    a_{\min}$,
  \item $\epsilon'_i < a_{\min}$,  $i=2,3,\ldots,n$,
  \item $\epsilon''_i \leq \epsilon_i$, $i=2,3,\ldots,n$,
  \end{enumerate}
  where $\epsilon'_i$ and $\epsilon''_i$ are obtained from
  $\epsilon_{i-1}$ using the formulas given in (\ref{e:e'}) and
  (\ref{e:e''}) with $a\geq a_{\min}$, respectively.
\end{lemma}
\begin{IEEEproof}
  Part 1) is obvious from the construction.  From (\ref{e:e'}) and 1),
  we have $\epsilon'_2 \leq \epsilon'_3 \leq \ldots \leq
  \epsilon'_n$. Thus it suffices to show $\epsilon'_n < a_{\min}$ when
  establishing 2). Indeed, note that $\epsilon_{n-1} <
  \frac{a_{\min}\tilde a_{n-1}}{2} \epsilon_s$, which together with
  (\ref{e:e'}) imply
  \begin{align*}
    \epsilon'_n &\leq \frac{1}{(a_{\min}-\epsilon_s)\tilde a_{n-1}} \Bigg\{
    a_{\min}^2 \left[\epsilon_s + a_{\min} \left(\frac{\tilde a_{n-1}
          \epsilon_s}{2} \right)^{n-1}\right] \\
    & ~~~~~~~~~~~~~~~~~~~~~ + (n-1)(1+\epsilon_s) \frac{\tilde a_{n-1}^2
      \epsilon_s}{2} \Bigg\} \\
    &\leq \frac{\epsilon_s}{(a_{\min}-\epsilon_s)\tilde a_{n-1}} \Bigg\{
    a_{\min}^2 \left[ 1 + \left(\frac{a_{\min} \tilde a_{n-1}}{2}
      \right)^{n-1}\right] \\
    & ~~~~~~~~~~~~~~~~~~~~~ + (n-1)(1+a_{\min})\frac{\tilde a_{n-1}^2}{2} \Bigg\} \\
    & < a_{\min}
  \end{align*}
  where the last two inequalities result from the upper bound imposed
  on $\epsilon_s$ in the statement of the lemma.
 
  Finally, it is easy to see from (\ref{e:e''}) that, for $i=2,3,\ldots,n$,
  \[
  \epsilon''_i \leq \frac{1}{\tilde a_{n-1} \epsilon_s}
  (\epsilon_{i-1}+\epsilon) \leq \frac{2\epsilon_{i-1}}{\tilde a_{n-1}
    \epsilon_s} = \epsilon_{i}
  \]
  where the second inequality is due to 1).
\end{IEEEproof}
Inductively applying Lemmas~\ref{t3} and \ref{t2b} with the choice of
$\epsilon_n$ from Lemma~\ref{thm:recur}, we obtain the following
theorem:
\begin{theorem}\label{theorem1}
  Let $2\leq n \leq \abs{U}$.  Suppose that $\epsilon_s$, $\epsilon$,
  and $\epsilon_{1}, \epsilon_{2}, \ldots, \epsilon_{n}$ are chosen
  according to Lemma~\ref{thm:recur}, and that the left null space of
  $A$ contains a normalized, $(a_{\min},\epsilon)$-polarized, $n
  \times \abs{U}$ matrix. Then the left null space contains either a
  normalized, $(a_{\min}, \epsilon'_i)$-double polarized, $1
  \!\times\!  \abs{U}$ vector for some $i=2,3,\ldots,n$, or a
  normalized, $(a_{\min},\epsilon_{n})$-diagonal polarized, $n
  \!\times\! \abs{U} $ matrix. In the latter case, no row in the
  original $(a_{\min},\epsilon)$-polarized matrix can be
  $\epsilon_s$-dependent upon the other rows.
\end{theorem}
\begin{IEEEproof}
  Let $\Upsilon$ be the original $(a_{\min},\epsilon)$-polarized
  matrix in the left null space of $A$ given in the statement of the
  theorem. We will construct the desired $(a_{\min},
  \epsilon'_i)$-double polarized vector or
  $(a_{\min},\epsilon_n)$-diagonal polarized matrix by inductively
  applying Lemmas~\ref{t3} and \ref{t2b} to the rows of $\Upsilon$.
  
  First, set $\hat\Upsilon^{(1)}$ to be the first row of $\Upsilon$.
  The assumption of the theorem guarantees that $\hat\Upsilon^{(1)}$
  satisfies that requirement of being a normalized,
  $(a_{\min},\epsilon_{1})$-diagonal polarized, $1\!\times\!\abs{U}$
  matrix in the left null space of $A$. Inductively, suppose that we
  have constructed, from the first $(i-1)$ rows of $\Upsilon$, the
  $(i-1)\!\times\!\abs{U}$ normalized matrix $\hat\Upsilon^{(i-1)}$
  that is $(a_{\min},\epsilon_{i-1})$-diagonal polarized and in the
  left null space of $A$.  If the $i$th row of $\Upsilon$ is
  $\epsilon_{s}$-dependent on the rows of $\hat\Upsilon^{(i-1)}$
  (i.e., the first $(i-1)$ rows of $\Upsilon$), then by
  Lemmas~\ref{t3} and \ref{thm:recur}, there exists a normalized,
  $(a_{\min}, \epsilon'_i)$-double polarized (at $(m,i)$ with $m<i$),
  $1 \!\times\! \abs{U}$ vector in the left null space of $A$. In this
  case, the induction process terminates.  On the other hand, if the
  $i$th row of $\Upsilon$ is not $\epsilon_{s}$-dependent on the rows
  of $\Upsilon^{(i-1)}$, then Lemmas~\ref{t2b} and \ref{thm:recur}
  together give an normalized $i \!\times\!  \abs{U}$ matrix
  $\hat\Upsilon^{(i)}$ that is $(a_{\min}, \epsilon_{i})$-diagonal
  polarized and in the left null space of $A$. Also note that the rows
  of $\hat\Upsilon^{(i)}$ are in the span of the first $i$ rows of
  $\Upsilon$. The induction process continues until $i=n$.
\end{IEEEproof}
Theorem~\ref{theorem1} leads to the following result that is critical
to development in the next section:
\begin{cor}\label{thm:nodbpolar}
  Fix $2 \leq n \leq \abs{U}$, a positive $\epsilon_s$ satisfying
  (\ref{e:es}), and $\frac{a_{\min}^2
    \epsilon_s}{(a_{\min}-\epsilon_s) \tilde a_{n-1}} < \epsilon' <
  a_{\min}$.  Let the left null space of $A$ contain no normalized
  $(a_{\min},\epsilon')$-double polarized, $1 \!\times\!  \abs{U}$
  vector. Then there exists a positive $\epsilon$ that if the left
  null space of $A$ contains a normalized,
  $(a_{\min},\epsilon)$-polarized $n \!\times\! \abs{U}$ matrix, no
  row of such matrix can be $\epsilon_s$-dependent upon the other
  rows.
\end{cor}
\begin{IEEEproof}
  For $i=1,2,\ldots,n-1$, choose $\epsilon_i = \epsilon \left(
    \frac{2}{\tilde a_{n-1} \epsilon_s}\right)^{i-1}$ as in
  Lemma~\ref{thm:recur}. For any $\delta>0$, from (\ref{e:e'}), there
  exists a small enough $\epsilon$ such that $\epsilon'_n \leq
  \frac{a_{\min}^2 \epsilon_s}{(a_{\min}-\epsilon_s) \tilde a_{n-1}} +
  \delta$.  Choosing $\delta= \epsilon' - \frac{a_{\min}^2
    \epsilon_s}{(a_{\min}-\epsilon_s) \tilde a_{n-1}}$ and using part
  1) of Lemma~\ref{thm:recur} give us $\epsilon'_2 \leq \epsilon'_3
  \leq \ldots \leq \epsilon'_n \leq \epsilon'$. Now applying
  Theorem~\ref{theorem1} gives us the desired result.
\end{IEEEproof}

The next lemma states that the observation channel $(A,B)$ being
non-manipulable is sufficient for the condition of non-existence of
any normalized $(a_{\min},\epsilon')$-double polarized, $1 \!\times\!
\abs{U}$ vector in the left null space of $A$ required in
Corollary~\ref{thm:nodbpolar}:
\begin{lemma} \label{thm:nonm=nodbp}
If $(A,B)$ is non-manipulable, then  there exists a pair of 
  constants $\epsilon_s$ and $\epsilon_A$ respectively satisfying
\begin{align}
  &0< \epsilon_s < \min_{1\leq n\leq\abs{U}-1} a_{\min}^2 \tilde a_{n}
  \Bigg\{ a_{\min}^2  \left[ 1 + \left(\frac{a_{\min} \tilde
        a_{n}}{2} \right)^{n}\right] \nonumber \\
  & ~~~~~~~~~~~~~~~~~~~~~~~~+ n(1+a_{\min})\frac{\tilde a_{n}^2}{2} +
  a_{\min}\tilde a_{n} \Bigg\}^{-1},
  \label{e:esmax}
\\
 &\frac{a_{\min}^2 \epsilon_s}{(a_{\min}-\epsilon_s) \tilde
    a_{\abs{U}-1}} < \epsilon_A < a_{\min},
\label{e:e'max}
\end{align}
such that the left null space of $A$ does not contain any normalized,
$(a_{\min},\epsilon_A)$-double polarized vectors.
\end{lemma}
\begin{IEEEproof}
  First we claim that the left null space of $A$ can not contain any
  normalized, $(a_{\min},0)$-double polarized vector if $(A,B)$ is
  non-manipulable. Indeed, suppose on the contrary that $\omega$ is a
  normalized vector in the left null space of $A$ that is
  $(a_{\min},0)$-double polarized at $(\alpha,\beta)$. Construct the
  $\abs{U}\!\times\! 1$ column vector $(\Upsilon_{\alpha}^T)^T$ whose
  $\alpha$th element is $\omega_{\alpha}$, $\beta$th elements is
  $-\omega_{\alpha}$, and all other elements are zero. Similarly,
  construct the $\abs{U}\!\times\! 1$ column vector
  $(\Upsilon_{\beta}^T)^T$ whose $\alpha$th element is
  $\omega_{\beta}$, $\beta$th elements is $-\omega_{\beta}$, and all
  other elements are zero. Then it is easy to check that
  $(\Upsilon_{\alpha}^T)^T$ and $(\Upsilon_{\beta}^T)^T$ are
  $(a_{\min},0)$-polarized at $\alpha$ and at $\beta$,
  respectively. Further, we also have, for all
  $l=1,2,\ldots,\abs{X_1}$, $(\Upsilon_{\alpha}^T)^T A_{\alpha,l} +
  (\Upsilon_{\beta}^T)^T A_{\beta,l} = 0$ because $\omega$ is in the
  left null space of $A$ and $\omega_i = 0$ for all $i \neq \alpha$ or
  $\beta$. Hence $(A,B)$ is manipulable.

  Next notice that both the set of normalized, $(a_{\min},0)$-double
  polarized vectors and the left null space of $A$ are closed
  sets. The former set is also bounded.  As a result, if the left null
  space of $A$ does not contain any normalized, $(a_{\min},0)$-double
  polarized vectors, it must also not contain any normalized,
  $(a_{\min},\epsilon)$-double polarized vectors for all small enough
  positive $\epsilon$.  Therefore by choosing $\epsilon_s$, satisfying
  (\ref{e:esmax}), small enough, we obtain an $\epsilon_A$ that
  satisfies (\ref{e:e'max}) and that the left null space of $A$
  contains no $(a_{\min},\epsilon_A)$-double polarized vectors.
\end{IEEEproof}

We now turn our attention to the right null space of $B$.  The
following result states that normalized vectors in the right null
space of $B$ have similar properties of normalized vectors in the left
null space of $A$ as described in Lemma~\ref{thm:lumaxnull}:
\begin{lemma}\label{thm:lumbxnull} 
  The right null space of $B$ consists only of balanced vectors.
  Suppose that $\omega^T$ is a non-zero normalized $\abs{U}\!\times\!
  1$ vector in the right null space of $B$. Let
\[
  b_{\min} \defn \frac{1}{\abs{U} (\abs{Y_1} +1)} .
\]
Then 
\begin{align*}
\max_{j:\omega_j > 0} \omega_j &\geq  b_{\min} \\
\min_{j:\omega_j < 0} \omega_j &\leq -b_{\min}.
\end{align*}
\end{lemma}
\begin{IEEEproof}
  Let $\omega^T$ be a vector in the right null space of $B$. Then
\[
\sum_{i=1}^{\abs{Y_1}}\sum_{j=1}^{\abs{U}} B_{i,j} \omega_j = 0.
\]
Swapping the order of the two sums and using the fact that
$\sum_{i=1}^{\abs{Y_1}} B_{i,j}=1$, we get $\sum_{j=1}^{\abs{U}}
\omega_j = 0$. Furthermore suppose $\omega^T$ is non-zero and
normalized, it must then have at least one positive element and one
negative element. Recognizing the preceding fact, we can employ
essentially the same argument in the proof of
Lemma~\ref{thm:lumaxnull} to show $\max_{j:\omega_j > 0} \omega_j \geq
b_{\min}$ and $\min_{j:\omega_j < 0} \omega_j \leq -b_{\min}$.
\end{IEEEproof}
Based on Lemma~\ref{thm:lumbxnull}, it is easy to check that the
results from Lemma~\ref{t3} to Corollary~\ref{thm:nodbpolar} all apply
to $B^T$ with $a_{\min}$ replaced by $b_{\min}$ and $\tilde a_n$
defined in Lemma~\ref{thm:recur} replaced by $\tilde b_n \defn
b_{\min}^2 \left( 1 + \frac{n}{b_{\min}} \right)^{-1}$. That is, the
above results are all applicable to normalized, polarized vectors and
matrices in the right null space of $B$ with the corresponding
modifications.

Finally, the next theorem states that non-manipulability of the
observation channel $(A,B)$ guarantees the existence of a counterpart
of Corollary~\ref{thm:nodbpolar} for the right null space of $B$ that
is to be used in the proof of Theorem~\ref{th:bound}.  To simplify
notation in statement of the lemma, let $\mathcal{A}$ denote the set
of all $\abs{U}^2\!\times\!  1$ column vectors of the form
$\mathrm{vec}( \Upsilon A) $, where $\Upsilon$ is a
$\abs{U}\!\times\!\abs{U}$ matrix whose $j$th column, for
$j=1,2,\ldots,\abs{U}$, is balanced and $(0,0)$-polarized at $j$, and
$\eu{\Upsilon} \leq 2\abs{U}$.  It is easy to check that the set
$\mathcal{A}$ is closed, bounded, and convex. Let $\mathcal{\tilde A}$
denote the cone hull of $\mathcal{A}$. Then it can be readily checked
that $\mathcal{\tilde A}$ is the set of vectors of the same form that
make up $\mathcal{A}$ with the norm bound $\eu{\Upsilon} \leq 2
\abs{U}$ removed.
\begin{theorem}\label{thm:(B,A)}

  Suppose that the right null space of $B$ is non-trivial, and the
  observation channel $(A,B)$ is non-manipulable.  Then there exists a
  positive $\kappa$, which depends only on $A$ and $B$, satisfying the
  property that if $\Psi$ is a $\abs{U}\!\times\!\abs{U}$ matrix whose
  columns are vectors in the right null space of $B$ and $\labs{\Psi}
  = 1$, then there is a normalized vector $\upsilon$ simultaneously
  giving $\upsilon\,\mathrm{vec}(\Psi) \geq \kappa$ and
  $\upsilon\omega^T \leq 0$ for all $\omega^T \in \mathcal{A}$.
\end{theorem}
\begin{IEEEproof}
  Let $n=\abs{U} - \mathrm{rank}(B) \geq 1$ be the dimension of the
  right null space of $B$. Let $\Xi$ be a $\abs{U}\!\times\! n$ matrix
  whose columns form an orthonormal basis of the right null space of
  $B$. By flipping the polarities of the columns of $\Xi$ (i.e., the
  basis vectors), we obtain $2^n$ different bases for the right null
  space of $B$.  Fix a normalized $\Psi$ with columns in the right
  null space of $B$. It is simple to check that $\frac{1}{\abs{U}^2}
  \leq \eu{\Psi}^2 \leq 1$. For each $j=1,2,\ldots,\abs{U}$, employing
  one, say $\Xi(j)$, among the $2^n$ bases above we can decompose
  $\Psi^T_j$ as $\Psi^T_j = \sum_{i=1}^n \left\{ \Psi^T_j
    (\Xi(j)^T_i)^T \right\} \, \Xi(j)^T_i$ with $\Psi^T_j
  (\Xi(j)^T_i)^T \geq 0$ for all $i$. Let $\mathcal{B}$ be the convex
  hull of the set of vectors of the form $\mathrm{vec}(\Phi)$, where
  $\Phi$ is any $\abs{U}\!\times\!\abs{U}$ matrix such that $\Phi^T_j
  = \sum_{i=1}^n b_{i,j} \,\Xi(j)^T_i$ for $j=1,2,\ldots, \abs{U}$
  with $\frac{1}{\abs{U}^2} \leq \sum_{i=1}^n \sum_{j=1}^{\abs{U}}
  b_{i,j}^2 \leq 1$ and $b_{i,j} \geq 0$ for $i$ and $j$.  Obviously
  $\mathrm{vec}(\Psi) \in \mathcal{B}$. By geometric reasoning,
  $\mathcal{B}$ is a bounded set that does not contain the origin.

  Since $(A,B)$ is non-manipulable, $\mathcal{\tilde A}$ must
  intersect trivially with the set of vectors of the form
  $\mathrm{vec}(\tilde\Phi)$, where $\tilde\Phi$ is any
  $\abs{U}\!\times\!\abs{U}$ matrix whose columns are vectors in the
  right null space of $B$ (i.e., the intersection contains only the
  zero vector).  Hence $\mathcal{B}$ and $\mathcal{\tilde A}$ are
  disjoint.  Below we employ a slightly stronger version of the
  argument given in \cite[pp. 48]{boyd2004co} to show that
  $\mathcal{B}$ and $\mathcal{\tilde A}$ can be strictly separated by
  a hyperplane that passes through the origin.

  Given the above, we conclude that there exist $\mu_{a}^T \in
  \mathcal{\tilde A}$ and $\mu_b^T \in \mathcal{B}$ that achieve the
  minimum (positive) Euclidean distance between $\mathcal{\tilde A}$
  and $\mathcal{B}$. Now let $\mu_1^T$ be an arbitrary vector in
  $\mathcal{\tilde A}$. Since $\mathcal{\tilde A}$ is a convex
  cone, $\mu_a^T + t(\alpha \mu_1^T - \mu_a^T) \in \mathcal{\tilde
    A}$ for all $\alpha > 0$ and $0 < t \leq 1$. Hence $\eu{\mu_a +
    t(\alpha\mu_1 - \mu_a) - \mu_b}^2 \geq \eu{\mu_a - \mu_b}^2$ for
  all $\alpha > 0$ and $0 < t \leq 1$, which (by letting $t \downarrow
  0$) implies
  \begin{equation} \label{e:A(l)} 
    (\mu_a - \mu_b) \mu_1^T \geq
    \frac{1}{2\alpha} \left(\eu{\mu_a - \mu_b}^2 + \eu{\mu_a}^2 -
      \eu{\mu_b}^2 \right)
  \end{equation}
  for all $\alpha > 0$.  Further, letting $\alpha \rightarrow \infty$
  gives
  \begin{equation} \label{e:A(l)'}
  (\mu_a - \mu_b) \mu_1^T \geq 0.
  \end{equation}
  Similarly, if $\mu_2^T$ be an arbitrary vector in $\mathcal{B}$,
  then $\mu_b^T + t(\mu_2^T - \mu_b^T) \in \mathcal{B}$ by the
  convexity of $\mathcal{B}$. Then $\eu{\mu_b + t(\mu_2 - \mu_b) -
    \mu_a}^2 \geq \eu{\mu_a - \mu_b}^2$ for all $0 < t \leq 1$ gives
  \begin{align} \label{e:B} 
   (\mu_b - \mu_a) \mu_2^T & \geq
    \frac{1}{2} \left(\eu{\mu_a - \mu_b}^2 + \eu{\mu_b}^2 -
      \eu{\mu_a}^2 \right) \nonumber \\
    & \geq \eu{\mu_a - \mu_b}^2
  \end{align}
  where the second inequality is due to the fact that $\eu{\mu_b}^2 -
  \eu{\mu_a}^2 \geq \eu{\mu_a - \mu_b}^2$, which can in turn be
  verified by letting $\mu_1^T = 0$ in (\ref{e:A(l)}) since $0 \in
  \mathcal{A}$.  Substituting $\mu_1=\omega$,
  $\mu_2^T=\mathrm{vec}(\Psi)$,
  $\upsilon=\frac{\mu_b-\mu_a}{\labs{\mu_b-\mu_a}}$, and
  $\kappa_{\mathcal{B}} = \frac{\eu{\mu_b -
      \mu_a}^2}{\labs{\mu_b-\mu_a}}$ in (\ref{e:A(l)'}) and
  (\ref{e:B}) almost establishes the lemma. The only technicality left
  to handle is that $\kappa_{\mathcal{B}}$ depends on
  $\Psi$. Fortunately, the dependence on $\Psi$ is only through the
  basis collection $\Xi(1), \Xi(2), \ldots, \Xi(\abs{U})$ that
  produces $\mathcal{B}$. Since there are only finitely many such
  collections to start with, the theorem is established by letting the
  required $\kappa$ to be the minimum among all the $2^{n\abs{U}}$
  $\kappa_{\mathcal{B}}$'s for the corresponding basis collections.
\end{IEEEproof}

\subsection{Bounding attack channel estimation error} \label{sec:theorems}

Recall that the attack channel matrix $\Phi^N$ is a $\abs{U}
\!\times\!\abs{U}$ stochastic matrix.  We consider below the
stochastic matrix $\Gamma^N = B \Phi^N A$ as defined in
(\ref{e:gammaN}), as well as estimates $\hat\Phi$ and $\hat\Gamma$ of
$\Phi^N$ and $\Gamma^N$, respectively. In particular, we are
interested in the estimators of $\Phi^N$ in $G_{\mu}(\hat\Gamma)$
defined in Section~\ref{sec:attackdetect}. For the purpose of proving
Theorem~\ref{thm:main} in the next section, the following result,
which bounds
the estimation error of estimators in $G_{\mu}(\hat\Gamma)$, is
important:
\begin{theorem} \label{th:bound} 
  Let $\hat\Gamma$ be an estimate of $\Gamma^N$ based on the
  observation $(y_1^N,x_1^N)$.  Let $\mu>0$ and $\hat\Phi \in
  G_{\mu}(\hat\Gamma)$ be an estimate of $\Phi^N$. If $(A,B)$ is
  non-manipulable, then
\[
\labs{\Phi^N - \hat{\Phi}}
  \leq   c_1 \mu + 
  c_2 \,\labs{ \Gamma^N - \hat\Gamma} + c_3 \,\labs{\Phi^N - I} 
\]
for some positive constants $c_1$, $c_2$, and $c_3$ that depend only
on $A$ and $B$.
\end{theorem}

\begin{IEEEproof}
  Let $\Xi_A$ and $\Xi_B$ denote the orthogonal projectors onto the
  left null space of $A$ and right null space of $B$, respectively.
  We decompose $\Phi^N - \hat{\Phi}$ into three components as below:
\begin{align}
\Phi^N - \hat{\Phi} &= 
 \underbrace{(\Phi^N - \hat{\Phi}) \Xi_A}_{\Theta^A \Psi^A} + 
 \underbrace{\Xi_B (\Phi^N - \hat{\Phi}) (I - \Xi_A)}_{\Lambda^B}
 \nonumber \\
  & ~~~~~+ \underbrace{(I - \Xi_B) (\Phi^N - \hat{\Phi}) (I - \Xi_A)}_{\Lambda}
\label{e:decomp}
\end{align}
where the rows of $\Psi^A$ are normalized, and $\Theta^A$ is a
$\abs{U} \!\times\! \abs{U}$ diagonal matrices whose strictly positive
diagonal elements are the normalization constants for the rows of
$\Psi^A$.  Note that the rows of $\Psi^A$ are vectors in the left null
space of $A$.  Moreover the columns of $\Lambda^B$ are vectors in the
right null space of $B$.  Also note that we have assumed that $\Psi^A$
does not contain any all-zero rows without any loss of generality (see
(\ref{e:decompboundA}) below).  Because the rows of $\Psi^A$ are
normalized, we have from (\ref{e:decomp}),
\begin{equation}\label{e:decompboundA}
  \labs{\Phi^N - \hat{\Phi}} 
  \leq \labs{\Lambda^B + \Lambda}+ \sum_{i=1}^{\abs{U}} \Theta^A_{i,i}.
\end{equation}
Thus it suffices to bound $\labs{\Lambda^B + \Lambda}$ and the
diagonal elements of $\Theta^A$.

We first bound the diagonal elements of $\Theta^A$. To do this,
rewrite (\ref{e:decomp}) as
\begin{equation}
  I - \hat{\Phi} = I - \Phi^N  + \Theta^A \Psi^A +\Lambda^B + \Lambda.
\label{e:I-hPhi}
\end{equation}
Because $\hat{\Phi}$ is a valid stochastic matrix, all diagonal
elements of $I- \hat{\Phi}$ must be greater than or equal to $0$ and
all off-diagonal elements must be less than or equal $0$. Thus
(\ref{e:I-hPhi}) gives
\begin{equation}
 \Theta^A_{i,i} \Psi^A_{i,j} 
\begin{cases}
  \geq -\labs{I - \Phi^N} - \labs{\Lambda^B + \Lambda} &
  \mbox{~if~} j=i \\
  \leq \labs{I - \Phi^N} + \labs{\Lambda^B + \Lambda} &
  \mbox{~if~} j \neq i.
\end{cases}
\label{e:ThetaPsi}
\end{equation}
Now by Lemma~\ref{thm:nonm=nodbp}, we have $\epsilon_s$ and
$\epsilon_A$ respectively satisfy (\ref{e:esmax}) and (\ref{e:e'max})
such that the left null space of $A$ does not contain any
$(a_{\min},\epsilon_A)$-double polarized vectors.  Hence we can choose
a positive $\epsilon < a_{\min}$ so that the conclusion in
Corollary~\ref{thm:nodbpolar} is valid for all $n=2,3,\ldots,\abs{U}$.
For this $\epsilon$, define
\[
S_{\epsilon} = \{ i : \big\vert \Psi^A_{i} \text{ is }
(a_{\min},\epsilon)\text{-polarized at }i \}.
\]
Let its cardinality be denoted by $n_{\epsilon}$.  

We bound $\Theta^A_{i,i}$ for $i\notin S_{\epsilon}$ and $i\in
S_{\epsilon}$ separately.  First consider any $i \notin
S_{\epsilon}$. Lemma~\ref{thm:offdiag} states that $[\Psi^A_i]_j \geq
\epsilon$ for some $j\neq i$.  Thus we have from (\ref{e:ThetaPsi})
that
\begin{equation}\label{e:ThetainSe}
  \Theta^A_{i,i} \leq \frac{\labs{I - \Phi^N}+ \labs{\Lambda^B +
      \Lambda} }{\epsilon}, 
\end{equation}
for all $i \notin S_{\epsilon}$.  

Next we bound $\Theta^A_{i,i}$ for $i\in S_{\epsilon}$. If
$n_{\epsilon}=0$, then there is nothing to do. Hence we assume $1 \leq
n_{\epsilon} \leq \abs{U}$ below. Since each column of $\Lambda^B$ are
in the right null space of $B$, it must be balanced according to
Lemma~\ref{thm:lumbxnull}. Moreover it is true that each column of
$\Phi^N - \hat{\Phi}$ must also be balanced. As a result,
(\ref{e:decomp}) gives $\sum_{i=1}^{\abs{U}} \Theta^A_{i,i}
\Psi^A_{i,j} = -\sum_{i=1}^{\abs{U}} \Lambda_{i,j}$ for
$j=1,2,\ldots,\abs{U}$. The triangle inequality then implies
\begin{equation} \label{e:thetaAbd} 
\left|\sum_{i=1}^{\abs{U}} \Theta^A_{i,i} \Psi^A_{i,j} \right| \leq
\sum_{i=1}^{\abs{U}} \abs{\Lambda_{i,j}} .
\end{equation}
Separating the sum on the left side of (\ref{e:thetaAbd}) into terms
with index $i$ in and not in $S_{\epsilon}$, we have, for
$j=1,2,\ldots,\abs{U}$,
\begin{eqnarray}
  \lefteqn{ \left|\sum_{i \in S_{\epsilon}} \Theta^A_{i,i}
      \Psi^A_{i,j}\right|} \nonumber \\
&\leq&
  \sum_{i=1}^{\abs{U}} \abs{\Lambda_{i,j}} 
  + \sum_{i \notin S_{\epsilon}} \Theta^A_{i,i} \abs{\Psi^A_{i,j}}
  \nonumber \\
&\leq&
  \sum_{i=1}^{\abs{U}} \abs{\Lambda_{i,j}} +
   \frac{\abs{U}}{\epsilon} \left( \labs{I - \Phi^N} + 
     \labs{\Lambda^B + \Lambda} \right)
\label{eq:Sec}
\end{eqnarray}
where the second inequality is obtained by using (\ref{e:ThetainSe})
and the fact that the rows of $\Psi^A$ are normalized.  For $2 \leq
n_{\epsilon} \leq \abs{U}$, Corollary~\ref{thm:nodbpolar} suggests
that for each $i\in S_{\epsilon}$,
\begin{align}\label{eq:upperbound}
  \sum_{j=1}^{\abs{U}} \left| \sum_{k \in S_{\epsilon}} \Theta^A_{k,k}
    \Psi^A_{k,j} \right| & = \Theta^A_{i,i} \sum_{j=1}^{\abs{U}} \left|
    \Psi^A_{i,j} + \sum_{k \in S_{\epsilon}, k\neq i}
    \frac{\Theta^A_{k,k}}{\Theta^A_{i,i}}
    \Psi^A_{k,j} \right|  \nonumber \\
  &> \Theta^A_{i,i} \epsilon_{s}.
\end{align}
Note that the bound in (\ref{eq:upperbound}) is also trivially valid
for the case of $n_{\epsilon}=1$ since $\epsilon_s < 1$ and the rows
of $\Psi^A$ are normalized.  Substituting (\ref{eq:Sec}) into
(\ref{eq:upperbound}), we obtain that for each $i \in S_{\epsilon}$,
\begin{equation} \label{eq:theta} 
\Theta^A_{i,i} < \frac{\labs{\Lambda}}{\epsilon_s} +
  \frac{\abs{U}^2}{\epsilon\epsilon_s}
  \left( \labs{I - \Phi^N}+\labs{\Lambda^B + \Lambda}\right) 
\end{equation}
Because $\epsilon_{s} \leq 1$, the upper bound on $\Theta^A_{i,i}$ for
$i \in S_{\epsilon}$ is greater than the upper bound on
$\Theta^A_{i,i}$ for $i \notin S_{\epsilon}$. Thus we can employ
(\ref{eq:theta}) as an upper bound for all $\Theta^A_{i,i}$'s.
Note that the choices of $\epsilon_A$, $\epsilon_s$, and $\epsilon$
depend only $A$.

Next we proceed to bound $\labs{\Lambda^B}$.  Similar to before, write
$\Lambda^B A = \theta^B \Psi^B$, where $\labs{\Psi^B}=1$ and
$\theta^B$ is the non-negative scaling factor.  If the right null
space of $B$ is trivial, $\Lambda^B = 0$ and hence
$\theta^B=0$. Otherwise, $\theta^B > 0$. Without loss of generality,
suppose the latter is true below.  Because the linear mapping
$(\cdot)A$ with domain restricted to the orthogonal complement of the
left null space of $A$ is invertible, we have
\begin{equation} \label{e:LambdaBA}
  \labs{\Lambda^B} \leq c_A \labs{\Lambda^B A} = c_A \theta^B
\end{equation}
for some constant $c_A$ that depends only on $A$. Thus bounding
$\theta^B$ is sufficient. To that end, right
multiply both sides of (\ref{e:I-hPhi}) by $A$ to obtain
\begin{equation}
  \underbrace{(I - \hat{\Phi})A}_{\Delta} = \theta^B \Psi^B  + 
  (I - \Phi^N) A  + \Lambda A.
\label{e:(I-hPhi)A}
\end{equation}
Since $\hat\Phi$ is stochastic, $\mathrm{vec}(\Delta) \in
\mathcal{A}$, where $\mathcal{A}$ is the set of $\abs{U}^2\!\times\!
1$ column vectors defined just right before Theorem~\ref{thm:(B,A)} in
Section~\ref{sec:lemmas}.  As $(A,B)$ is non-manipulable and the
left-null space of $B$ is non-trivial, Theorem~\ref{thm:(B,A)}
guarantees the existence of a positive constant $\kappa$ and a
normalized vector $\nu$ giving $\nu \,\mathrm{vec}(\Delta) \leq 0$ and
$\nu \,\mathrm{vec}(\Psi^B) \geq \kappa$. Note that $\kappa$ depends
only on $A$ and $B$.  Hence left-multiplying both sides of the
vectorized version of (\ref{e:(I-hPhi)A}) by $\nu$ yields
\begin{align*}
\theta^B & \leq \frac{\left| \nu \, \{\mathrm{vec}((I - \Phi^N)A) +
  \mathrm{vec}(\Lambda A) \} \right| }{\kappa} \nonumber \\
& \leq \frac{\labs{(I - \Phi^N)A} + \labs{\Lambda A}} {\kappa}
\nonumber \\
& \leq \frac{\sqrt{\abs{U}\abs{X_1}^2}}{\kappa} \left( \labs{I -
    \Phi^N} + \labs{\Lambda} \right).
\end{align*}
Applying this bound on $\theta^B$ back to (\ref{e:LambdaBA}), we
obtain
\begin{equation} \label{e:thetaB} 
  \labs{\Lambda^B} \leq
  \frac{c_{A}\sqrt{\abs{U}\abs{X_1}^2}}{\kappa} \left( \labs{I -
      \Phi^N} + \labs{\Lambda} \right).
\end{equation}

To complete the proof, we need to bound $\labs{\Lambda}$. Note that
there exists $\tilde\Gamma$ such that $B \hat\Phi A = \Pi_B
\tilde\Gamma\Pi_A$ and $\labs{\Pi_B(\tilde\Gamma-\hat\Gamma)\Pi_A}
\leq \mu$ since $\hat\Phi \in G_{\mu}(\hat\Gamma)$. Then
\begin{align}
  B \Lambda A & = B (\Phi^N - \hat{\Phi})A = \Gamma^N - \Pi_B
  \tilde\Gamma\Pi_A
  \nonumber \\
  & = \Pi_B (\Gamma^N - \hat{\Gamma}) \Pi_A + \Pi_B (\hat\Gamma -
  \tilde\Gamma)\Pi_A
\label{e:BlambdaA}
\end{align}
where the last equality is due to the fact that $\Gamma^N = \Pi_B
\Gamma^N \Pi_A$.  Note that the linear mapping $B(\cdot)A$ with domain
restricted simultaneously to the orthogonal complements of the right
null space of $B$ and left null space of $A$ is invertible.  Combining
this fact and (\ref{e:BlambdaA}), there exists a positive constant
$c_{AB}$, which depends only on $A$ and $B$, such that
\begin{equation}\label{eq:lambda}
  \labs{\Lambda} \leq c_{AB} \left(\sqrt{\abs{X_1}\abs{Y_1}}
    \cdot\labs{\Gamma^N - \hat{\Gamma}}+\mu\right).
\end{equation}

Finally, substituting (\ref{eq:lambda}), (\ref{e:thetaB}), and
(\ref{eq:theta}) back into (\ref{e:decompboundA}),
we obtain the desired bound on $\labs{\Phi^N-\hat\Phi}$ given in the
statement of the theorem with
\begin{align*}
  c_1 & = \frac{c_{AB}\abs{U}}{\epsilon_{s}} + c_{AB}
  \left( 1 + \frac{\abs{U}^3}{\epsilon\epsilon_s} \right)
  \left( 1 + \frac{c_{A}\sqrt{\abs{U}\abs{X_1}^2}}{\kappa} \right)
 \\
 c_2 & = \sqrt{\abs{X_1}\abs{Y_1}} \cdot c_1 \\
 c_3 & = \frac{\abs{U}^3}{\epsilon\epsilon_s} +
  \frac{c_{A}\sqrt{\abs{U}\abs{X_1}^2}}{\kappa} 
  \left( 1 + \frac{\abs{U}^3}{\epsilon\epsilon_s} \right).
\end{align*}
\end{IEEEproof}

\subsection{Proof of Theorem~\ref{thm:main} and
  Corollary~\ref{thm:maincor}} \label{sec:proofmain}

\subsubsection{Detectability}
We prove the detectability portion of the theorem by showing that the
sequence of decision statistics $\{D^N = \labs{\hat\Phi^N - I}\}$,
where $\hat{\Phi}^N$ is the estimator for the stochastic matrix
$\Phi^N$ defined in (\ref{e:maxhatPhi}), satisfies the two desired
properties if the observation channel $(A,B)$ is non-manipulable.

To that end, first note that $\hat{\Phi}^N$ is obtained from the
conditional histogram estimator $\hat{\Gamma}^N$ of the stochastic
matrix $\Gamma^{N}$ defined in (\ref{e:histo}). We need the following
convergence property of the sequence $\{\hat{\Gamma}^N\}$:
\begin{lemma}\label{thm:^GammaN}
  $\labs{{\Gamma}^N - \hat{\Gamma}^N} \rightarrow 0$ in probability as
  $N$ approaches infinity.
\end{lemma}
\begin{IEEEproof}
First, define the $\abs{U}\!\times\!\abs{X_1}$ stochastic matrix
$\Omega^N$ by its $(i,j)$th element as
\begin{equation}\label{e:OmegaN}
\Omega^N_{i,j} \defn \frac{\pi^N(v_i,x_{1,j})}{\pi^N(x_{1,j})}.
\end{equation}
For any $\mu>0$, it is clear that
\begin{eqnarray}
\lefteqn{\Pr\left( \labs{{\Gamma}^N - \hat{\Gamma}^N} > \mu \right)}
\nonumber \\
&\leq& \hspace*{-8pt} 
\Pr\left( \labs{B\Omega^N - \hat{\Gamma}^N} > \frac{\mu}{2} \right)
+
\Pr\left( \labs{B\Omega^N - {\Gamma}^N} > \frac{\mu}{2} \right)
\nonumber \\
&\leq& \hspace*{-8pt} 
\Pr\left( \labs{B\Omega^N - \hat{\Gamma}^N} > \frac{\mu}{2} \right)
\nonumber \\
& & + \Pr\left(\labs{\Omega^N - \Phi^N A} >
\frac{\mu}{2\sqrt{\abs{X_1}\abs{Y_1}} \cdot\eu{B}} \right) .
\label{e:omegabound}
\end{eqnarray}
Thus the lemma is proved if we can show that the two probabilities on
the right hand side of (\ref{e:omegabound}) converge to $0$ as $N$
approaches infinity.

To that end, notice first that
\begin{eqnarray*}
  \lefteqn{\labs{B\Omega^N - \hat{\Gamma}^N} } \\
  &\hspace*{-9pt}=&  \hspace*{-6pt}
  \sum_{i=1}^{\abs{Y_1}} \sum_{j=1}^{\abs{X_1}}
  \abs{[B\Omega^N]_{i,j} - \hat{\Gamma}^N_{i,j}}  \\
   &\hspace*{-9pt}\leq& \hspace*{-6pt}
  \sum_{i=1}^{\abs{Y_1}} \sum_{j=1}^{\abs{X_1}} \sum_{k=1}^{\abs{U}} 
  \Big| \underbrace{p(y_{1,i}|v_k) 
    \frac{\pi^N(v_k,x_{1,j})}{\pi^N(x_{1,j})}
    -\frac{\pi^N(y_{1,i},v_k,x_{1,j})}{\pi^N(x_{1,j})}}_{H_{i,j,k}}
  \Big|
\end{eqnarray*}
where the inequality above is due to the fact that
$\pi^N(y_{1,i},x_{1,j}) = \sum_k \pi^N(y_{1,i},v_k,x_{1,j})$. This
implies that
\begin{eqnarray}
\lefteqn{\Pr\left( \labs{B\Omega^N - \hat{\Gamma}^N} > \frac{\mu}{2}
  \right)} \nonumber \\
  &\leq& 
  \sum_{i=1}^{\abs{Y_1}} \sum_{j=1}^{\abs{X_1}} \sum_{k=1}^{\abs{U}} 
  \Pr\left(|H_{i,j,k}| > \frac{\mu}{2\abs{U}\abs{X_1}\abs{Y_1}}\right).
  \label{e:omegabound1}
\end{eqnarray}
But for any $i$, $j$, and $k$,
\begin{eqnarray}
  \lefteqn{\Pr\left(|H_{i,j,k}| >
      \frac{\mu}{2\abs{U}\abs{X_1}\abs{Y_1}}\right)
    \leq 
    \frac{4\abs{U}^2\abs{X_1}^2\abs{Y_1}^2}{\mu^2} 
    E \left[ H_{i,j,k}^2 \right] } \nonumber \\
  &\leq &  \hspace*{-8pt}
  \frac{4\abs{U}^2\abs{X_1}^2\abs{Y_1}^2}{\mu^2} \left\{
    \Pr(x^N_1 \notin T^N_{[X_1],\delta}) + E_{[X_1],\delta} [ H_{i,j,k}^2 ] \right\},
  \nonumber \\
\label{e:PrH}
\end{eqnarray}
where $E_{[X_1],\delta} [\cdot]$ denotes that conditional expectation
$E[\cdot|x^N_1\in T^N_{[X_1],\delta}]$. Further, for any small
enough positive $\delta$,
\begin{eqnarray}
  \lefteqn{E_{[X_1],\delta}[ H_{i,j,k}^2]} \nonumber \\
  &\leq& \hspace*{-8pt}
  \frac{ E_{[X_1],\delta} \left[ (p(y_{1,i}|v_k) \pi^N(v_k,x_{1,j})
      -\pi^N(y_{1,i},v_k,x_{1,j}))^2 \right]}{N^2(p(x_{1,j})-\delta)^2} 
  \nonumber \\
  &=& \hspace*{-8pt}
  \frac{ \sum_{n=1}^{N} E_{[X_1],\delta} \left[
   \begin{array}{l}
      E \left[(p(y_{1,i}|v_k)-1_n(y_{1,i}))^2|v^N,x_1^N \right] \\
      ~\cdot 1_n(v_k,x_{1,j})
   \end{array}
    \right]}{N^2(p(x_{1,j})-\delta)^2} 
  \nonumber \\
  &\leq& \hspace*{-8pt}
  \frac{p(y_{1,i}|v_k)}{N(p(x_{1,j})-\delta)^2}
\label{e:EHbound}
\end{eqnarray}
where the equality on the third line above results from the fact that
$p(y^N_1|v^N,x^N_1) = p(y^N_1|v^N)$ and the elements of $y^N_1$ are
conditionally independent given $v^N$. Combining (\ref{e:EHbound}) and
the well known fact, for example see \cite[Theorem 6.2]{Yeung2008},
that $\Pr(x^N_1 \notin T^N_{[X_1],\delta}) \rightarrow 0$ as $N
\rightarrow \infty$, we get from (\ref{e:PrH}) that
$\Pr\left(|H_{i,j,k}| > \frac{\mu}{2\abs{U}\abs{X_1}\abs{Y_1}}\right)
\rightarrow 0$ as $N \rightarrow \infty$. Using (\ref{e:omegabound1}),
we further get $\Pr\left( \labs{B\Omega^N - \hat{\Gamma}^N} >
  \frac{\mu}{2} \right) \rightarrow 0$ as $N \rightarrow \infty$.

Next, note that we can rewrite (\ref{e:OmegaN}) as
\begin{align*}
\Omega^{N}_{i,j} &= \frac{\pi^N(u_{i} , x_{1,j})}{\pi^N(x_{1,j})}
- \sum_{k \neq i} \frac{\pi^N(v_{k} ,u_{i} ,x_{1,j})}{\pi^N(x_{1,j})} \\
& ~~~~+ \sum_{k \neq i} \frac{\pi^N(v_{i} ,u_{k},x_{1,j})}{\pi^N(x_{1,j})} .
\end{align*}
Similarly, we have
\begin{align*}
\left[\Phi^{N}A \right]_{i,j} &= p(u_{i} \vert x_{1,j}) - \sum_{k \neq i}
\frac{\pi^N(v_{k} , u_{i})}{\pi^N(u_i)} p(u_{i} \vert x_{1,j}) \\
& ~~~~
+ \sum_{k\neq i} \frac{\pi^N(v_{i} , u_{k})}{\pi^N(u_{k})} p(u_{k} \vert
x_{j}) .
\end{align*}
Let 
\begin{align*}
H^1_{i,j} &= \frac{\pi^N(u_{i}, x_{1,j})}{\pi^N(x_{1,j})}-p(u_{i}\vert x_{1,j})
\\ 
H^{2}_{i,j} &= \sum_{k \neq i} \frac{\pi^N(v_{k} , u_{i} , x_{1,j})}{\pi^N(x_{1,j})} - \frac{\pi^N(v_{k} , u_{i})}{ \pi^N(u_{i} )} p(u_{i} \vert x_{1,j})
\\
H^{3}_{i,j} &= \sum_{k \neq i} \frac{\pi^N(v_{i} , u_{k} ,
  x_{1,j})}{\pi^N(x_{1,j})} - \frac{\pi^N(v_{i} , u_{k})}{ \pi^N(u_{k} )}
p(u_{k} \vert x_{1,j}).
\end{align*}
Then we have, for each $i$ and $j$,
\begin{eqnarray}
\lefteqn{\Pr\left(\left|\Omega^N_{i,j} - [\Phi^N A]_{i,j} \right| >
    \frac{\mu}{2\abs{U}\abs{X_1}\sqrt{\abs{X_1}\abs{Y_1}}
        \cdot\eu{B}} \right)} \nonumber \\
 & \leq & \hspace*{-8pt}
 \sum_{l=1}^{3}
 \Pr\left(\abs{H^l_{i,j}}>\frac{\mu}{6\abs{U}\abs{X_1}\sqrt{\abs{X_1}\abs{Y_1}}
     \cdot\eu{B}}\right).
\label{e:omega-phiA}
\end{eqnarray}
By employing the fact that $p(x^N_1|u^N,v^N)=p(x^N_1|u^N)$ and the
conditional independence of the elements of $u^N$ given $x^N_1$, each
of the three probabilities on the right hand side of
(\ref{e:omega-phiA}) can be shown to converge to $0$ as $N$ approaches
infinity by using typicality arguments similar to the one that shows
$\Pr\left(|H_{i,j,k}| > \frac{\mu}{2\abs{U}\abs{X_1}\abs{Y_1}}\right)
\rightarrow 0$ above. Thus, the probability on the left hand side of
(\ref{e:omega-phiA}) also converges to $0$ as $N$ approaches
infinity. Finally,
\begin{eqnarray*}
\lefteqn{\Pr\left(\labs{\Omega^N - \Phi^N A} >
\frac{\mu}{2\sqrt{\abs{X_1}\abs{Y_1}} \cdot\eu{B}} \right)} \\
& \leq & 
\sum_{i=1}^{\abs{U}} \sum_{j=1}^{\abs{X_1}}
\Pr\Bigg(\left|\Omega^N_{i,j} - [\Phi^N A]_{i,j} \right| \\
& & ~~~~~~~~~~~~~~~~~~>
    \frac{\mu}{2\abs{U}\abs{X_1}\sqrt{\abs{X_1}\abs{Y_1}}
        \cdot\eu{B}} \Bigg) \\
& \rightarrow & 0
\end{eqnarray*}
as $N \rightarrow \infty$.
\end{IEEEproof}

Now we proceed to show detectability in Theorem~\ref{thm:main}.  For
any fixed $\mu>0$, choose $\hat\Phi^{N}$ according to
(\ref{e:maxhatPhi}).  It is clear from the definition of
$G_{\mu}(\hat\Gamma^N)$ in Section~\ref{sec:attackdetect} that $\Phi^N
\in G_{\mu}(\hat\Gamma^N)$ whenever $\labs{\Gamma^N - \hat\Gamma^N}
\leq \frac{\mu}{\sqrt{\abs{X_1}\abs{Y_1}}}$. Thus
Lemma~\ref{thm:^GammaN} implies that $\Pr ( \Phi^{N} \in
G_{\mu}(\hat\Gamma^N) ) \rightarrow 1$ as $N \rightarrow \infty$, and
hence the probability that $G_{\mu}(\hat\Gamma^N)$ is non-empty
approaches $1$. We employ this property below to show that the
sequence of decision statistics $\{\labs{\hat\Phi^N-I}\}$ satisfies
both Properties~1 and~2 stated in the theorem.

To show Property~1 of Theorem~\ref{thm:main}, first note that
\begin{eqnarray}
\lefteqn{\Pr\left(\labs{\hat\Phi^{N} - I} >
  \delta~\bigcap~\labs{\Phi^{N} - I} >
  \delta\right)} \nonumber \\
&\geq& 
\Pr \Big( \Phi^{N} \in G_{\mu}(\hat\Gamma^{N})~\bigcap~
  \labs{\hat\Phi^{N} - I} > \delta \nonumber \\
& & ~~~~~~~\bigcap~\labs{\Phi^{N} - I} > \delta \Big) \nonumber \\
&=&
\Pr \left( \Phi^{N} \in G_{\mu}(\hat\Gamma^{N})~\bigcap~
  \labs{\Phi^{N} - I} > \delta \right)
\nonumber \\
&\geq&
  \Pr(\labs{\Phi^{N} - I} > \delta) - 
  \Pr(\Phi^{N} \notin G_{\mu}(\hat\Gamma^{N})) 
\label{e:condP1}
\end{eqnarray}
where the equality on the third line is due to the definition of
$\hat\Phi^{N}$ in (\ref{e:maxhatPhi}). Hence, if
$\limsup_{N\rightarrow\infty} \Pr(\labs{\Phi^{N} - I} > \delta) >0$,
(\ref{e:condP1}) implies
\[
\limsup_{N\rightarrow\infty} \Pr(\labs{\hat\Phi^{N} - I} >
\delta~|~\labs{\Phi^{N} - I} > \delta) = 1.
\]

To show Property~2 of Theorem~\ref{thm:main}, consider the constants
$c_1$, $c_2$, and $c_3$ given in Theorem~\ref{th:bound}. Let $c =
2+c_3$. By choosing $0<\mu \leq\frac{\delta}{2c_1}$ and making use of
Theorem~\ref{th:bound}, it is easy to check that
\begin{eqnarray*}
\lefteqn{\Pr \left(\labs{\Gamma^N - \hat\Gamma^{N}} \leq
    \frac{\mu}{\sqrt{\abs{X_1}\abs{Y_1}}} 
  ~\bigcap~\labs{\Phi^{N} - I} \leq 
  \delta \right)} \nonumber \\
&~~\leq& \hspace*{-6pt}
\Pr \left(\labs{\hat\Phi^{N} - I} \leq c\delta~\bigcap~\labs{\Phi^{N} - I} \leq 
  \delta \right)
\end{eqnarray*}
which in turn implies
\begin{eqnarray}
\lefteqn{\Pr \left(\labs{\hat\Phi^{N} - I} >  
  c\delta~\bigcap~\labs{\Phi^{N} - I} \leq 
  \delta \right)} \nonumber \\
&\leq& 
  \Pr \left(\labs{\Gamma^N - \hat\Gamma^{N}} >
    \frac{\mu}{\sqrt{\abs{X_1}\abs{Y_1}}} \right).~~~~~~~~~~~~~
\label{e:condP2}
\end{eqnarray}
Hence by Lemma~\ref{thm:^GammaN}, if $\liminf_{N\rightarrow\infty}
\Pr(\labs{\Phi^{N} - I} \leq \delta) >0$, (\ref{e:condP2}) gives
\[
\lim_{N\rightarrow\infty} \Pr(\labs{\hat\Phi^{N} - I} >
c\delta~|~\labs{\Phi^{N} - I} \leq \delta) = 0.
\]

\subsubsection{Corollary~\ref{thm:maincor}}
Now we can use the results above to prove Corollary~\ref{thm:maincor}
by showing the sequence of decision statistics
$\{\labs{\hat\Phi^N-I}\}$ satisfies the four stated properties under
the corresponding special cases:
\paragraph{$\labs{\Phi^N - I} \rightarrow 0$ in probability} Since
\begin{eqnarray*}
  \lefteqn{\Pr(\labs{\hat\Phi^{N} - I} >  
    \delta) } \\
  &\leq& 
  \Pr \left(\labs{\hat\Phi^{N} - I} >  
    \delta~\Big|~\labs{\Phi^{N} - I} \leq 
    \frac{\delta}{c} \right) \\
  & & ~\cdot \Pr \left(\labs{\Phi^{N} - I} \leq 
    \frac{\delta}{c} \right) + \Pr \left(\labs{\Phi^{N} - I} > 
    \frac{\delta}{c} \right)
\end{eqnarray*}
for any $\delta>0$, Property~2 of Theorem~\ref{thm:main} implies
$\lim_{N\rightarrow\infty} \Pr(\labs{\hat\Phi^{N} - I} > \delta) = 0$.

\paragraph{$\labs{\Phi^N - I} \nrightarrow 0$ in probability}
If $\limsup_{N\rightarrow\infty} \Pr(\labs{\Phi^{N} - I} > \delta) =
0$, then there is nothing to show. Otherwise, Lemma~\ref{thm:^GammaN}
and~(\ref{e:condP1}) give $\limsup_{N\rightarrow\infty}
\Pr(\labs{\hat\Phi^{N} - I} > \delta) \geq
\limsup_{N\rightarrow\infty} \Pr(\labs{\Phi^{N} - I} > \delta)$.

\paragraph{ $\labs{\Phi^{N_M} - \Phi} \rightarrow 0$ in
  probability and $\Phi \neq I$}
Note that there exists $\delta>0$ such that
$\limsup_{N\rightarrow\infty} \Pr(\labs{\Phi^{N} - I} > \delta) = 1$
in this case. For any such $\delta$, Property~2 just above implies the
required result.

\paragraph{$\labs{\Phi^N - \Phi} \rightarrow 0$ in probability and
  $\Phi \neq I$}
In this case, there exists $\delta>0$ such that
$\lim_{N\rightarrow\infty} \Pr(\labs{\Phi^{N} - I} > \delta) =
1$. Using Lemma~\ref{thm:^GammaN} and~(\ref{e:condP1}) gives the
desired result for any such $\delta$.

\subsubsection{Converse}

Recall that $x_{1}^N$ and $y_1^N$ are the sequences of symbols
transmitted and received by node 1 at time instants $1,2,\ldots,N$ and
$N+1,N+2,\ldots,2N$, respectively.  Let $D^N = D^N(x_1^N,y_1^N)$ be
the $N$th decision statistic in the sequence, assumed to exist in the
statement of Theorem~\ref{thm:main}, that satisfies Properties~1 and
2. Suppose that there exists a stochastic $\Phi' \neq I$ such that $B
\Phi' A = BA \defn \Gamma$.

For the identity relay manipulation map, i.e., $\phi^N(u^N)=u^N$, it
is easy to check that $p(v^n|u^n) = \prod_{n=1}^N I_{\chi_n(v^N),
  \chi_n(u^N)}$, $\Phi^N = I$, and $p(y^N|x^N) = \prod_{n=1}^N
\Gamma_{\chi_n(y_1^N), \chi_n(x_1^N)}$. Since $\Phi^N = I$ for all
$N$, we have for every $\delta >0$,
\begin{eqnarray}
  \lefteqn{ \hspace*{-30pt}
    \sum_{(x_{1}^{N},y_{1}^{N}) : D^N(x_{1}^{N},y_{1}^{N})> \delta} 
    \underbrace{\left[ \prod_{i=n}^{N}\Gamma_{\chi_n(y_1^N), \chi_n(x_1^N)} \cdot
        p(x_{1,\chi_n(x_1^N)}) \right]}_{q(y_1^N,x_1^N)}  } \nonumber \\
  &=& 
  \Pr (D^{N}  > \delta) \nonumber \\
  &=&
  \Pr \left( D^{N} > \delta~\Big|~  \labs{\Phi^{N} - I} \leq 
    \frac{\delta}{c} \right) \nonumber \\
  &\rightarrow& 0
\label{eq:nm}
\end{eqnarray}
where the convergence on the last line is due to the assumption that
Property~2 of Theorem~\ref{thm:main} holds, and $c$ is the constant
described in the property.

Consider now the random relay manipulation map that results in
$p(v^n|u^n) = \prod_{n=1}^N \Phi'_{\chi_n(v^N), \chi_n(u^N)}$. For
this random manipulation map, it is again easy to check that
$p(y^N|x^N) = \prod_{n=1}^N \Gamma_{\chi_n(y_1^N), \chi_n(x_1^N)}$,
which is a consequence of the assumption that $B \Phi' A = \Gamma$,
and $\labs{\Phi^N - \Phi'} \rightarrow 0$ in probability. Since $\Phi'
\neq I$, there exists a $\delta>0$ such that $\lim_{N\rightarrow
  \infty} \Pr ( \labs{\Phi^N - I} > \delta) = 1$. For this $\delta$,
\begin{eqnarray}
  \lefteqn{\limsup_{N\rightarrow \infty} 
    \sum_{(x_{1}^{N},y_{1}^{N}) : D^N(x_{1}^{N},y_{1}^{N}) > \delta} 
    q(y_1^N, x_1^N)  } \nonumber \\
  &=& 
  \limsup_{N\rightarrow \infty} 
  \Pr ( D^N > \delta) \nonumber \\
  &\geq&
  \limsup_{N\rightarrow \infty} 
  \Pr \left( D^{N} >
    \delta~\Big|~ \labs{\Phi^{N} - I} > \delta \right) \nonumber \\
  & & ~\cdot 
  \lim_{N\rightarrow \infty} 
  \Pr ( \labs{\Phi^{N} - I} > \delta)
  \nonumber \\
  &=& 1
\label{eq:m2}
\end{eqnarray}
where the equality on the last line is due to the assumption that
Property~1 of Theorem~\ref{thm:main} holds.  The conclusions in
(\ref{eq:nm}) and (\ref{eq:m2}) are clearly in conflict and can not be
simultaneously true. Therefore there can not exist a stochastic
$\Phi' \neq I$ such that $B\Phi' A = BA $ and Properties~1 and~2 hold.

To complete the proof of the converse, we need to show the existence
of a stochastic $\Phi' \neq I$ with the property that $B \Phi' A = BA$
when the observation channel $(A,B)$ is manipulable. To that end,
first note that the manipulability of $(A,B)$ implies the existence of
a non-zero $\Upsilon$ with the property that all columns of $\Upsilon
A$ are in the right null space of $B$. In addition, $(\Upsilon^T_j)^T$
is balanced and $(0,0)$-polarized at $j$, for each
$j=1,2,\ldots,\abs{U}$. Let $\tilde \Upsilon = \frac{\Upsilon}{\max_j
  \Upsilon_{j,j}}$ and $\Phi' = I - \tilde\Upsilon$. It then easy to
check that this $\Phi'$ is a valid stochastic matrix, $\Phi' \neq 1$,
and $B\Phi' A = BA$.

\subsection{Justification for Algorithms~\ref{thm:check_mani} and
  \ref{thm:Aalg}}

\subsubsection*{Algorithm~\ref{thm:check_mani}}
Let $\Upsilon$ be a $\abs{U}\!\times\!\abs{U}$ matrix-valued variable
and $s$ be a $\abs{U}\!\times\! 1$ vector-valued variable. Consider
the following convex optimization problem:
\begin{align}
  &\min_{s,\Upsilon} ~ \min \{-s_1,-s_2,\ldots,-s_{\abs{U}}\} & \nonumber \\
  & \mathrm{\; subject\ to} & \nonumber \\
  & \quad B \Upsilon A = 0  \nonumber \\
  & \quad \sum_{k=1}^{\abs{U}} \Upsilon_{k,l} =  0 &
     l=1,2,\ldots,\abs{U},  \nonumber \\
  & \quad s_{k} -  \Upsilon_{k,k} \leq  0 &
     k=1,2,\ldots,\abs{U},  \nonumber \\
  & \quad \Upsilon_{k,k} - 1 \leq  0 &
     k=1,2,\ldots,\abs{U},  \nonumber \\
  & \quad \Upsilon_{k,l} \leq 0 & k\neq
  l=1,2,\ldots,\abs{U}.
  \label{e:conv_opt}
\end{align}
It is clear that the optimal value of (\ref{e:conv_opt}) is attained
and lies inside the interval $[-1,0]$. This further implies that
$(A,B)$ is non-manipulable if and only if the optimal value of
(\ref{e:conv_opt}) is $0$.

Now, let $\Omega$ and $\nu$ be the $\abs{X_1}\!\times\!\abs{Y_1}$
matrix-valued and $\abs{U}\!\times\! 1$ vector-valued Lagrange
multipliers for the equality constraints shown in the third and fourth
lines of (\ref{e:conv_opt}).  Further, let $\Lambda$ be the
$\abs{U}\!\times\!\abs{U}$ matrix-valued Lagrange multiplier
matrix. The diagonal elements of $\Lambda$ correspond to the
inequality constraints shown in the fifth line of (\ref{e:conv_opt}),
while the off-diagonal elements correspond to the inequality
constraints shown in the last line of (\ref{e:conv_opt}). At last, let
$\nu$ be the $\abs{U}\!\times\! 1$ vector-valued Lagrange multiplier
for the inequality constraints shown in second line of
(\ref{e:conv_opt}).  Following the development in
\cite[Ch. 5]{boyd2004co}, we obtain the Lagrange dual function of
(\ref{e:conv_opt}) as below:
\begin{align*}
&g(\gamma,\Lambda,\nu,\Omega)  \\
& = 
\begin{cases}
  - \sum_{k=1}^{\abs{U}} \gamma_k &   \hspace*{-20pt}\text{if}\\
  \quad ~\Lambda_{k,k} \geq 1 & k=1,2,\ldots,\abs{U}, \\
  \quad ~\gamma_k + \nu_k + [A\Omega B]_{k,k} = \Lambda_{k,k} &
  k=1,2,\ldots,\abs{U},  \\
  \quad ~\nu_k + [A\Omega B]_{k,l} = - \Lambda_{k,l} & k\neq
  l=1,2,\ldots,\abs{U}
  \\
  -\infty & \hspace*{-20pt}\text{otherwise.}
  \end{cases}
\end{align*}
Consider the Lagrange dual problem of (\ref{e:conv_opt}):
\begin{align}
  &\max_{\gamma,\Lambda,\nu,\Omega} ~g(\gamma,\Lambda,\nu,\Omega)  & \nonumber \\
  & \mathrm{\; subject\ to} & \nonumber \\
  & \quad \gamma_k \geq 0  & k=1,2,\ldots,\abs{U}, \nonumber \\
  & \quad \Lambda_{k,l} \geq 0  & k,l=1,2,\ldots,\abs{U}.
  \label{e:dual}
\end{align}
Since the primal problem (\ref{e:conv_opt}) satisfies the Slater's
condition \cite[Ch. 5]{boyd2004co}, the optimal duality gap between
the primal and dual problems is zero. That is, the optimal value of
(\ref{e:dual}) is the same as the optimal value of (\ref{e:conv_opt}).
Hence $(A,B)$ is non-manipulable if and only if the optimal value of
(\ref{e:dual}) is $0$. Finally, it is easy to verify that the
negation of the optimal value of the linear program in
Step~\ref{step:opt} of Algorithm~\ref{thm:check_mani} is the optimal
value of (\ref{e:dual}) and vice versa. As a result,
Algorithm~\ref{thm:check_mani} determines whether $(A,B)$ is
manipulable.

\subsubsection*{Proof of Theorem~\ref{thm:B=I}}
In the proof of Lemma~\ref{thm:nonm=nodbp}, we have shown that the
condition of $(A,B)$ being non-manipulable implies that no normalized,
$(a_{\min},0)$-double polarized vector can be in the left null space
of $A$. It remains to show the reverse implication here.

Given that the left null space of $A$ does not contain any
$(a_{\min},0)$-double polarized vector, we need to show that $(A,B)$
is non-manipulable when the right null space of $B$ is trivial. To
that end, let us suppose on the contrary that $(A,B)$ is
manipulable. Since the right null space of $B$ is trivial, there
exists a $\abs{U}\!\times\!\abs{U}$ non-zero matrix $\Upsilon$ in the
left null space of $A$, with its $j$th column, $(\Upsilon^T_j)^T$, for
each $j=1,2,\ldots,\abs{U}$, is balanced and $(0,0)$-polarized at $j$.
Let $m$ be the number of non-zero columns of $\Upsilon$.  By proper
permutation of rows and columns of $\Upsilon$ if necessary, we can
assume with no loss of generality that the first $m$ columns are
non-zero while the remaining $\abs{U}-m$ columns are zero. 

First, we claim that $m \geq 2$. Indeed, suppose that $m=1$ and only
the first column $(\Upsilon_1^T)^T$ is non-zero. Then we have
$\Upsilon_{1,1} > 0$ and $\Upsilon_{1,1} A_{1} = 0$. Since it is our
assumption that $A$ contains no zero rows, we must have
$\Upsilon_{1,1}=0$, which creates a contradiction.

Next, let 
\begin{align*}
  \tilde\Upsilon_1 &= \frac{\Upsilon_1}{\labs{\Upsilon_1}} \\
  \tilde\Upsilon_2 &= \frac{\Upsilon_2}{\labs{\Upsilon_2}} \\
  & \vdots \\
  \tilde\Upsilon_m &= \frac{\sum_{i=m}^{\abs{U}}
    \Upsilon_i}{\labs{\sum_{i=m}^{\abs{U}} \Upsilon_i}}.
\end{align*}
Put these $m$ row vectors together to form the $m\!\times\!\abs{U}$
matrix $\tilde \Upsilon$.  Using the fact that $(\Upsilon_j^T)^T$ is
balanced and $(0,0)$-polarized at $j$ for $j=1,2,\ldots,m$ together
with Lemma~\ref{thm:lumaxnull}, it is easy to check that $\tilde
\Upsilon$ is a normalized, $(a_{\min},0)$-polarized matrix in the left
null space of $A$. Moreover, it is also true that
\begin{equation}\label{e:lindep}
\sum_{i=1}^{m-1} \labs{\Upsilon_i} \,\tilde\Upsilon_i +
\bigg\|\sum_{i=m}^{\abs{U}} \Upsilon_i\bigg\|_1 \,\tilde\Upsilon_m = 0.
\end{equation}

Now, as argued in the proof of Lemma~\ref{thm:nonm=nodbp}, there must
exist $\epsilon_s$ and $\epsilon_A$, which respectively satisfy
(\ref{e:esmax}) and (\ref{e:e'max}), such that the left null space of
$A$ contains no $(a_{\min},\epsilon_A)$-double polarized vector. Hence
we can apply Corollary~\ref{thm:nodbpolar} to $\tilde\Upsilon$ to
deduce that no row of it can be $\epsilon_s$-dependent upon the other
rows. However, this conclusion contradicts (\ref{e:lindep}). Therefore
$(A,B)$ must not be manipulable.

\subsubsection*{Algorithm~\ref{thm:Aalg}}
We use Lemma~\ref{thm:lumaxnull} to show that the
Algorithm~\ref{thm:Aalg} can be used to check for double polarized
vectors in the left null space of $A$.  To that end, recall that
$n=\abs{U} - \textrm{rank}(A)$ is the dimension of the left null space
of $A$.  If $n=0$, the left null space of $A$ is trivial and hence it
can not contain any normalized, $(a_{\min},0)$-double polarized
vector. For $n>0$, let $\Upsilon = (I~\tilde\Upsilon)$ be the
row-reduced echelon basis matrix as stated in Step \ref{step:n=U-1})
of the algorithm. When $n=\abs{U}-1$, $\tilde\Upsilon_i$ is a scalar,
for $i=1,2,\ldots,n$. By Lemma~\ref{thm:lumaxnull}, $\tilde\Upsilon_i$
must be negative and the normalized version of $\Upsilon_i$ must be a
$(a_{\min},0)$-double polarized vector in the left null space of $A$.

Now assume $1 \leq n \leq \abs{U}-2$ and consider the stated
steps. First note that any $(a_{\min},0)$-double polarized vector in
the left null space of $A$ can only be a linear combination of at most
two rows of $\Upsilon$.  If in Step \ref{step:allzero}) there is a
$\tilde\Upsilon_i$ that contains all but one zero element,
Lemma~\ref{thm:lumaxnull} again forces the non-zero element be
negative and the normalized version of $\Upsilon_i$ be a
$(a_{\min},0)$-double polarized vector in the left null space of
$A$. Otherwise no (normalized) rows of $\Upsilon$ can be
$(a_{\min},0)$-double polarized. Hence it remains to check whether any
pair of rows of $\Upsilon$ can be linearly combined to form a double
polarized vector. Without loss of generality, suppose that the
normalized version of $c_1\Upsilon_1 + c_2\Upsilon_2$ is
$(a_{\min},0)$-double polarized for some non-zero constants $c_1$ and
$c_2$. Then it is easy to see that $c_1$ and $c_2$ must be of opposite
signs and $\tilde\Upsilon_1 = -
\frac{c_2}{c_1}\tilde\Upsilon_2$. Hence if no pairs of rows of
$\tilde\Upsilon$ satisfy the condition checked in Step \ref{step:=}),
the left null space of $A$ can not contain any normalized,
$(a_{\min},0)$-double polarized vector.

\section{Conclusions} \label{sec:conclusion} 

We showed that it is possible to detect whether an amplify-and-forward
relay is maliciously manipulating the symbols that it forwards to the
other nodes by just monitoring the relayed symbols.  In particular, we
established a non-manipulable condition on the channel that serves as
a necessary and sufficient requirement guaranteeing the existence of a
sequence of decision statistics that can be used to distinguish a
malicious relay from a non-malicious one.

An important conclusion of the result is that maliciousness
detectability in the context of Theorem~\ref{thm:main} is solely
determined by the source distributions and the conditional pmfs of the
underlying MAC and BC in the channel model, regardless of how the
relay may manipulate the symbols. Thus similar to capacity,
maliciousness detectability is in fact a channel characteristic. The
development of maliciousness detectability in this paper did not take
any restrictions on the rate and coding structure of information
transfer between the sources into account. A joint formulation of
maliciousness detectability and information transfer is currently
under investigation.

Another interesting application of the result is that the necessity of
non-manipulability can be employed to show that maliciousness
detectability is impossible for non amplify-and-forward relays in many
channel scenarios. For instance, if Romeo in the motivating example
considered in Section~\ref{se:motivate} forwards his received symbol
modulo-$2$ instead, we arrive at the physical-layer network coding
(PNC) model considered in \cite{Zhang2006}. The necessity condition of
Theorem~\ref{thm:main} can be employed to verify the impossibility of
maliciousness detectability with the PNC operation represented by the
matrix $B$. For this relaying operation, additional signaling may be
needed to allow for maliciousness detection.

\end{document}